\pdfoutput=1
\documentclass[sigconf,screen]{acmart}

\AtBeginDocument{%
  }

\newcommand{\DOI}{\href{https://doi.org/10.5281/zenodo.7493145}{https://doi.org/10.5281/zenodo.7493145 }}
\newcommand{\github}{\href{https://github.com/ampersand-projects/streambench}{https://github.com/ampersand-projects/streambench }}

\setcopyright{rightsretained}
\acmPrice{}
\acmDOI{10.1145/3575693.3575704}
\acmYear{2023}
\copyrightyear{2023}
\acmSubmissionID{asplosb23main-p65-p}
\acmISBN{978-1-4503-9916-6/23/03}
\acmConference[ASPLOS '23]{Proceedings of the 28th ACM International Conference on Architectural Support for Programming Languages and Operating Systems, Volume 2}{March 25--29, 2023}{Vancouver, BC, Canada}
\acmBooktitle{Proceedings of the 28th ACM International Conference on Architectural Support for Programming Languages and Operating Systems, Volume 2 (ASPLOS '23), March 25--29, 2023, Vancouver, BC, Canada}
\received{2022-07-07}
\received[accepted]{2022-09-22}

\usepackage[normalem]{ulem}
\usepackage{todonotes}
\usepackage{listings}
\usepackage{caption}
\usepackage{tabularx}
\usepackage[ruled,vlined]{algorithm2e}
\usepackage{hyperref}
\usepackage{microtype}
\usepackage{enumitem}
\usepackage{amsmath}
\usepackage{multirow}
\usepackage{subfig}
\usepackage{graphicx}
\usepackage{microtype}
\usepackage{tikz}
\usepackage{flushend}

\newcommand{\tilt}[1]{$\sim$${#1}$}

\begin{document}

\title[TiLT: A Time-Centric Approach for Stream Query Optimization and Parallelization]{TiLT: A Time-Centric Approach for\\ Stream Query Optimization and Parallelization}

\author{Anand Jayarajan}
\email{anandj@cs.toronto.edu}
\affiliation{%
  \institution{University of Toronto, Vector Institute}
  \country{Canada}
}

\author{Wei Zhao}
\authornote{Authors have contributed equally to this research.}
\email{xiaoteemo.zhao@mail.utoronto.ca}
\affiliation{%
  \institution{University of Toronto}
  \country{Canada}
}

\author{Yudi Sun}
\authornotemark[1]
\email{yudi@cs.toronto.edu}
\affiliation{%
  \institution{University of Toronto}
  \country{Canada}
}

\author{Gennady Pekhimenko}
\email{pekhimenko@cs.toronto.edu}
\affiliation{%
  \institution{University of Toronto, Vector Institute}
  \country{Canada}
}


\begin{abstract}
Stream processing engines (SPEs) are widely used for large scale streaming analytics over unbounded time-ordered data streams. Modern day streaming analytics applications exhibit diverse compute characteristics and demand strict latency and throughput requirements. Over the years, there has been significant attention in building hardware-efficient stream processing engines (SPEs) that support several query optimization, parallelization, and execution strategies to meet the performance requirements of large scale streaming analytics applications. However, in this work, we observe that these strategies often fail to generalize well on many real-world streaming analytics applications due to several inherent design limitations of current SPEs. We further argue that these limitations stem from the shortcomings of the fundamental design choices and the query representation model followed in modern SPEs. To address these challenges, we first propose \emph{TiLT}, a novel intermediate representation (IR) that offers a highly expressive temporal query language amenable to effective query optimization and parallelization strategies. We subsequently build a compiler backend for TiLT that applies such optimizations on streaming queries and generates hardware-efficient code to achieve high performance on multi-core stream query executions. We demonstrate that TiLT achieves up to $326\times$ ($20.49\times$ on average) higher throughput compared to state-of-the-art SPEs (e.g., Trill) across eight real-world streaming analytics applications. TiLT source code is available at \href{https://github.com/ampersand-projects/tilt.git}{https://github.com/ampersand-projects/tilt.git}.
\end{abstract}



\maketitle

\section{Introduction}
Stream processing applications are widely used in the industry to perform both real-time and offline analysis on unbounded, time-ordered, and high-frequency event streams. For example, social media companies (e.g., Twitter, Meta) use click stream analytics for serving advertisements~\cite{heron, twitter}, banking institutions analyze purchasing trends for identifying fraudulent transactions~\cite{fraud}, and investment companies conduct high-frequency trading based on real-time stock prices~\cite{hft}. In recent years, stream processing is finding even wider application in non-traditional areas like agriculture~\cite{farmbeat}, climate science~\cite{weather}, energy/manufacturing industry~\cite{vibration}, and healthcare~\cite{lifestream}. These applications involve computation that require \emph{fine-grained control over the time-dimension} of data streams, such as computing a moving average over time or finding the temporal correlation between events across streams. Moreover, streaming computations are often long-running and process data in a continuous manner with strict latency and throughput requirements~\cite{req}.

\begin{table}[]
    \centering
    \caption{Throughput (million events/sec) of Yahoo streaming benchmark~\cite{streambench} on a 32-core machine}
    \begin{tabular}{cc|cccc}
        \hline
        \multicolumn{2}{c|}{Scale-out\cite{spark,flink}} & \multicolumn{4}{c}{Scale-up~\cite{trill,streambox,grizzly,lightsaber}}\\
        \hline
        Spark          & Flink         & Trill & StreamBox & Grizzly & LightSaber \\ \hline
        0.14           & 0.59          & 34.07 & 167.19    & 118.74  & 296.40     \\ \hline
    \end{tabular}
    \label{tbl:ysb}
    \vspace{-10pt}
\end{table}

Stream processing engines (SPEs) are special systems designed to meet the ever-increasing performance demands of streaming analytics applications. Modern SPEs~\cite{spark, struct, flink, dataflow, trill, lifestream, streamql, grizzly, lightsaber, saber} provide familiar SQL-like temporal query language interface for writing complex streaming analytics applications. Many popular SPEs~\cite{spark, flink, dataflow, struct, storm} are designed as scale-out systems which meet the performance requirements of streaming applications by scaling the query execution over large cluster of machines. Despite being a widely adopted design, scale-out SPEs showcase significantly low hardware-utilization and therefore are highly resource intensive~\cite{analyze, scalecost}. This spawned several scale-up SPEs~\cite{trill, streambox, streamboxhbm, lightsaber, grizzly, lifestream} with hardware-conscious designs to better utilize modern multi-core machines. Table~\ref{tbl:ysb} shows the performance comparison between state-of-the-art scale-out and scale-up SPEs on the popular Yahoo streaming benchmark~\cite{streambench}. As shown, scale-up SPEs are able to achieve $100-1000\times$ higher throughput under the same hardware budget~\cite{analyze}, which makes them a more cost-effective alternative to scale-out SPEs for large scale stream processing~\cite{dowe}.

Scale-up SPEs face three key challenges to achieve high single-machine performance for streaming queries. First, improving hardware utilization of streaming queries requires supporting \emph{effective optimization strategies} for improving cache utilization of query execution and pruning redundant computation. Second, unlike batch processing applications, streaming queries do not inherently expose data parallelism due to their continuous query execution model. Therefore, SPEs need to support \emph{sophisticated parallelization strategies} to fully utilize all the processing cores in a multi-core machine. Finally, since stream queries are long-running workloads, they require a \emph{low-overhead runtime} to meet their latency requirements. Despite having significant research attention, current scale-up SPE designs fail to address all three challenges at the same time on many real-world streaming applications due to the following reasons.

Current scale-up SPEs~\cite{trill, streambox, streamboxhbm, brisk} follow the traditional data flow representation of streaming applications where the queries are defined as a directed acyclic graph (DAG) of temporal operations and the query execution is performed by interpreting the data flow graph. Even though this design is conceptually simple and easy to extend, this query execution model is shown to introduce significant interpretation overhead at runtime~\cite{analyze, neumann}. Moreover, the query optimizations in the interpreted SPEs are mostly heuristics-based graph-level transformations such as reordering the operations in the DAG~\cite{struct}. Applying such optimizations requires the streaming query to precisely match with certain pre-defined rules and, therefore, typically has narrow applicability on many streaming applications. Finally, many of these systems only extract limited parallelization opportunities available through partitioned data streams.

To address the inefficiencies of interpretation-based SPEs, recent works have proposed compiler-based solutions~\cite{saber, lightsaber, grizzly, scabbard}. These approaches offer low-overhead runtime for query execution by automatically generating hardware-efficient code from the high-level query description. State-of-the-art compiler-based SPEs also support low-level query optimizations like operator fusion to maximize data locality by passing data between operators through registers or cache memory, and can automatically parallelize the query execution even on non-partitioned data streams. However, the optimization, parallelization, and code generation strategies proposed in current compiler-based SPEs primarily target applications performing only a limited set of operations (e.g., stream aggregations) and these strategies do not generalize well on queries with more complex operations (e.g., stream-to-stream join). This significantly limits the ability of current compiler-based SPEs to support many real-world streaming analytics applications.

In this work, to support the growing adoption of stream processing in a wide range of application domains, we set a goal to provide an infrastructure for effective and generalizable optimization and parallelization strategies for streaming queries. We make a key observation that the limited optimization and parallelization capabilities of SPEs are due to the fundamental limitations of the query representation model used by modern SPEs. Under the current so-called \emph{event-centric} model, streaming queries are constructed using primitive temporal operations, each defining a transformation over a sequence of discrete time-ordered events. Even though this is a natural representation model for streaming queries as the data streams are inherently an unbounded time-ordered sequence of events, we argue that this event-centric definition of temporal operations does not expose the important time semantical information of the streaming queries that are required for effective query optimization and parallelization. Since streaming queries are temporal in nature, we believe that the temporal operations should also follow a representation model that is fundamentally based on time.

Based on this observation, we propose a novel intermediate representation (IR) called \emph{TiLT} that follows a \emph{time-centric} model for defining streaming queries. Unlike the traditional event-centric model, TiLT IR defines temporal operations as functional transformations over well-defined time-domains using new constructs like \emph{temporal object}, \emph{reduction function}, and \emph{temporal expression}. With these simple constructs, TiLT offers a highly expressive programming paradigm to represent a diverse set of streaming applications. At the same time, the time-centric definition of streaming queries enables optimization opportunities that are otherwise difficult to perform using the traditional query representation models. Moreover, the side-effect-free functional definition of TiLT queries exposes inherent data parallelism that can be leveraged to parallelize arbitrary streaming queries. Finally, we build a compiler-backend for TiLT that automatically translates the logical stream query definitions into hardware-efficient executable code and achieves high multi-core performance on a wide range of streaming applications.

To evaluate TiLT's ability to provide high performance on a diverse range of applications, we prepare a benchmark suite with eight stream processing applications representative of real-world streaming analytics use-cases in fields including stock trading, signal processing, industrial manufacturing, banking institutions, and healthcare. On these applications, TiLT achieves $6-322\times$ ($20.49\times$ on average) higher throughput against the state-of-the-art interpretation-based SPE Trill~\cite{trill}. This speedup comes from two major fronts: (i) effective query optimization and parallelization enabled by the time-centric query representation model in TiLT, and (ii) a compiler-based SPE design that eliminates common inefficiencies like query interpretation and managed language overhead common in interpreted SPEs. We also show that TiLT can achieve competitive performance against compiler-based SPEs that are specially designed for efficient stream aggregation. For example, on Yahoo streaming benchmark~\cite{benchstream}, TiLT is able to achieve $1.5\times$ and $3.8\times$ higher throughput compared to state-of-the-art compiler-based SPEs LightSaber~\cite{lightsaber} and Grizzly~\cite{grizzly}, respectively.

In summary, we make the following contributions:
\begin{itemize}
    \item We highlight the limitations of the query representation models used in current SPEs in supporting effective optimization and parallelization strategies. To address these limitations, we propose a novel intermediate representation called TiLT. We show that TiLT enables generalizable optimization and parallelization strategies that are otherwise difficult to support in the traditional query representation models.
    \item We build a compiler for TiLT that can optimize and parallelize arbitrary streaming queries and generate hardware-efficient code to achieve high multi-core performance.
    \item We prepare a new representative benchmark suite with eight real-world streaming analytics applications used in signal processing, stock trading, industrial manufacturing, banking service, and healthcare. Across these applications, we demonstrate that TiLT can achieve $6-326\times$ ($20.49\times$ on average) higher throughput compared to state-of-the-art SPEs. TiLT is currently open-sourced and available at \href{https://github.com/ampersand-projects/tilt.git}{https://github.com/ampersand-projects/tilt.git}
\end{itemize}
\pdfoutput=1
\section{Background}\label{sec:back}
\begin{table*}[]
    \centering
    \small
    \caption{Real-world streaming analytics applications}
    \begin{tabular}{|l|l|l|l|}
    \hline
    \multicolumn{1}{|c|}{\textbf{Analytics application}} & \multicolumn{1}{c|}{\textbf{Description}}                                                                                                 & \multicolumn{1}{c|}{\textbf{Operators in the query}}                        & \multicolumn{1}{c|}{\textbf{Data set}}                                                                                                    \\ \hline
    Trend-based trading~\cite{invest}                                  & Moving average trend in stock price                                                                                                       & Avg (2), Join, Where                                                        & \multirow{2}{*}{New York Stock Exchange~\cite{nyse}}                                                                                                  \\ \cline{1-3}
    Relative strength index~\cite{rsi}                        & Stock price momentum indicator                                                                                                            & Shift, Join, Avg (2)                                                        &                                                                                                                                           \\ \hline
    Normalization~\cite{norm}                                        & Normalize event values using Z-score                                                                                                      & Avg, StdDev, Join                                                           & \multirow{3}{*}{\begin{tabular}[c]{@{}l@{}}Synthetic Data (Random\\ floating point values generated\\ at $1000$Hz frequency)\end{tabular}} \\ \cline{1-3}
    Signal imputation~\cite{impute}                                    & Replacing missing signal values                                                                                                           & Avg, Shift, Join                                                            &                                                                                                                                           \\ \cline{1-3}
    Resampling~\cite{resample}                                           & Changing signal frequency                                                                                                                 & Select, Join, Shift, Chop                                                   &                                                                                                                                           \\ \hline
    Pan-Tomkins algorithm~\cite{pantom}                                & Detect QRS complexes in ECG                                                                                                               & Custom-Agg(3), Select, Avg                                                              & MIMIC-III waveform data~\cite{mimic}                                                                                                                   \\ \hline
    Vibration analysis~\cite{kurt}                                   & \begin{tabular}[c]{@{}l@{}}Monitor machine vibrations using kurtosis,\\ root mean square, and crest factor metrics\end{tabular} & \begin{tabular}[c]{@{}l@{}}Max, Avg(2), Join (2),\\ Custom-Agg\end{tabular} & Bearing vibration data~\cite{bear_data}                                                                                                                    \\ \hline
    Fraud detection~\cite{fraud}                                      & Credit card fraud detection                                                                                                               & Avg, StdDev, Shift, Join                                                    & Kaggle credit card data~\cite{qty_data}                                                                                                                   \\ \hline
    \end{tabular}
    \label{tbl:apps}
\end{table*}

Streaming analytics applications typically process unbounded sequence of time-ordered events in a continuous manner. Table~\ref{tbl:apps} shows eight representative real-world streaming analytics applications used in areas including stock market trading, signal processing, healthcare, manufacturing, and banking services. These applications process data at a high rate and demand strict latency and throughput requirements. For example, high-frequency trading applications demand sub-second level latency~\cite{lowlat, lowlat2, lowlat3}. Moreover, the computations performed by these applications often require \emph{fine-grained control over the time-dimension} of the data streams such as using different windowing strategies to analyze changing trends in the data streams~\cite{invest, rsi, kurt}.

\begin{figure*}
\centering
    \centering
    \includegraphics[width=\textwidth]{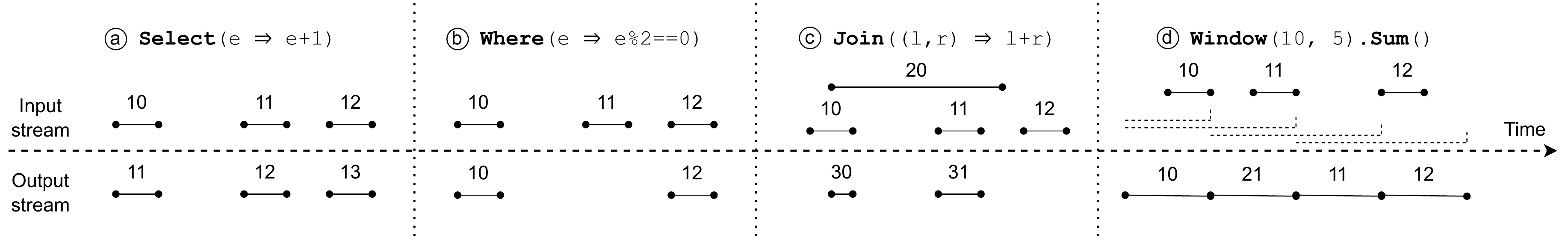}
    \caption{Common temporal operations: (a) Select, (b) Where, (c) Temporal Join, and (d) Window-Sum}
    \label{fig:ops}
\end{figure*}

Many modern SPEs~\cite{spark, flink, trill, lightsaber, grizzly, lifestream} provide SQL-like query languages with temporal extensions for writing such complex streaming applications. These languages offer a vocabulary of simple yet highly expressive primitive temporal operations with each defining a transformation over one or more data streams. Figure~\ref{fig:ops} illustrates four primitive temporal operations commonly used in streaming queries and their corresponding transformations on event streams. Without loss of generality, each event in the data stream is represented using a payload value and a validity interval. The \emph{Select} and \emph{Where} operations shown in the Figure~\ref{fig:ops}\textcolor{red}{a} and \ref{fig:ops}\textcolor{red}{b} follow the relational SQL semantics of projection and selection operations, respectively. Both operations perform per-event transformations where the former modifies the payload field of each event and the latter conditionally filters out events based on a user-defined predicate on the payload. The temporal \emph{Join} operation shown in Figure~\ref{fig:ops}\textcolor{red}{c} joins two streams into a single output stream. The output stream of the \emph{Join} operation contains events corresponding to the strictly overlapping regions of events in the input streams. Finally, the aggregation operations on data streams are generally performed over a time-bounded window defined by its window size and stride length. For example, a \emph{Sum} aggregation operation defined over a \emph{Window(size, stride)} computes the sum of every $size$-seconds\footnote{For the purpose of the discussion, we use seconds as the unit of time. However, any other units of time are also applicable to the definitions used in this paper.} windows that are \emph{stride}-seconds apart. Figure~\ref{fig:ops}\textcolor{red}{d} shows a sliding-window aggregation with window size $10$ and stride length $5$. Since all these operations are defined as transformations over events, we call them to follow an \emph{event-centric} model of temporal operator definition.

\begin{figure}
    \centering
    \subfloat[Un-optimized version]{
        \includegraphics[width=0.47\columnwidth]{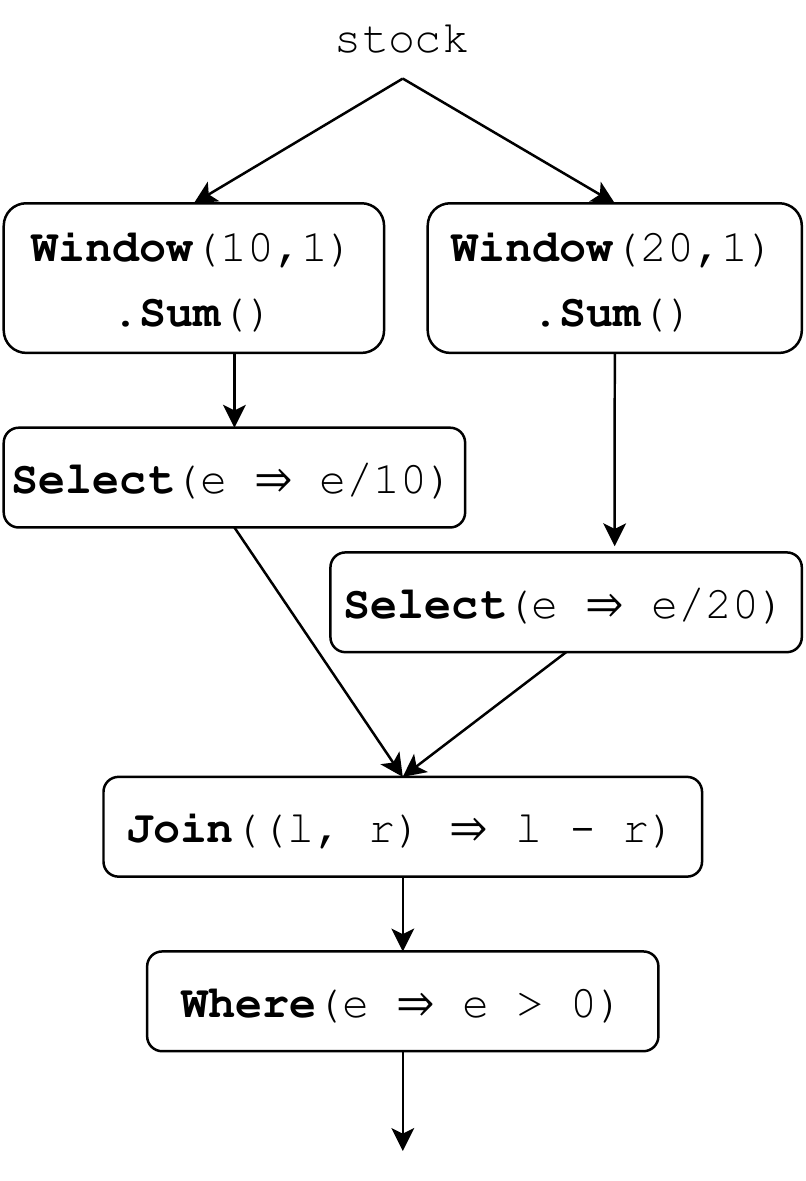}
        \label{fig:q}
    }
    \subfloat[Fused version]{
        \includegraphics[width=0.47\columnwidth]{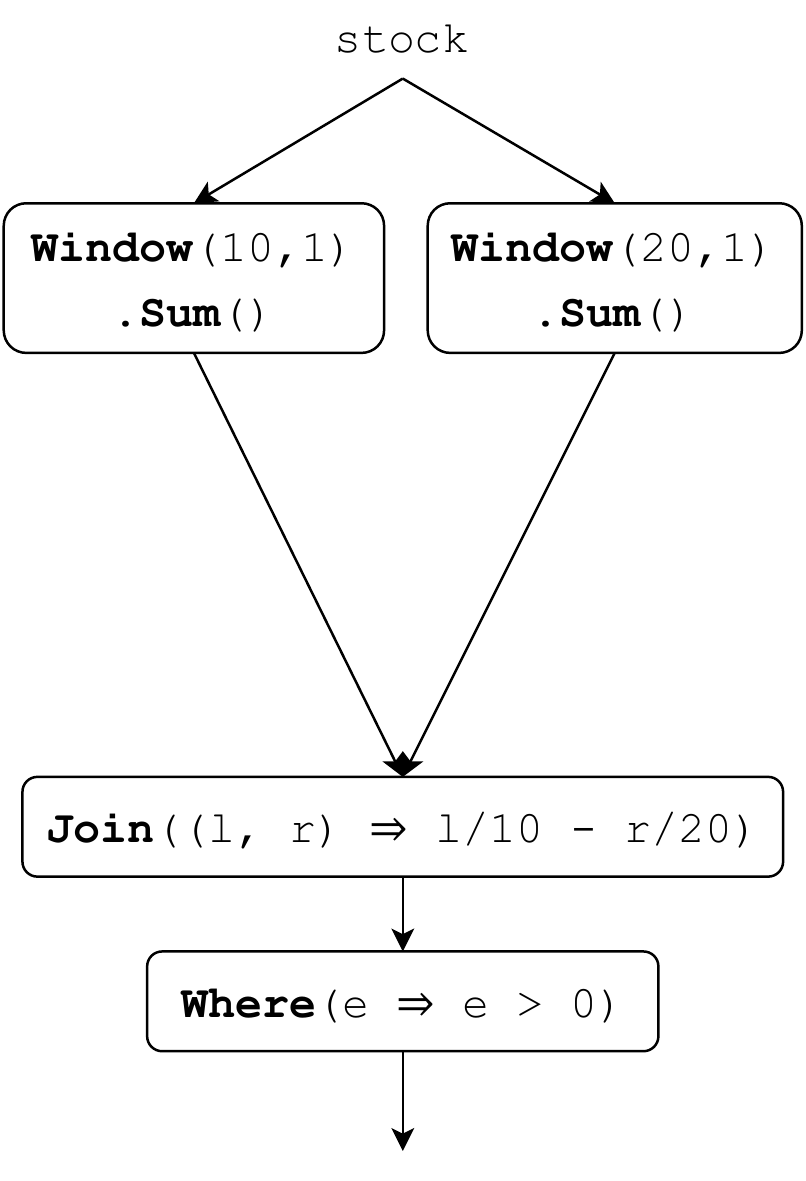}
        \label{fig:fuseq}
    }
    \caption{Stock price trend analysis query}
\end{figure}

These simple primitive operations can be combined together to construct more complex streaming queries. Figure~\ref{fig:q} shows an example streaming query written using the primitive temporal operations shown in Figure~\ref{fig:ops}. This query is a simplified version of the stock market trend analysis application~\cite{invest} described in Table~\ref{tbl:apps}. This query analyzes the trends in the price of a particular stock by computing two moving average of the stock prices over $10$ and $20$ seconds intervals on every second by first using two \emph{Window-Sum} operations and then dividing the sum by their corresponding window sizes using the \emph{Select} operation. Afterwards, the difference between the concurrent pairs of $10$-second and $20$-second averages is computed using the temporal \emph{Join} operation. The final \emph{Where} operation filters only the events with a positive difference in the average. The validity intervals of the events in the final output stream corresponds to the period of time for which the stock price is observing an upward trend.

Once the temporal queries are defined, SPEs can internally handle the execution of the queries. Many popular SPEs~\cite{spark, struct, flink, dataflow, millwheel, storm} are primarily designed to meet the performance requirements of the query by distributing the execution over large cluster of machines. Despite being a widely popular approach, prior works~\cite{trill, lifestream, analyze, terse} have shown that such scale-out SPEs are highly resource-intensive and often perform $2-3$ orders of magnitude slower than their corresponding hand-tuned implementations, thus causing significant waste of compute and energy resources. This observation led to the development of several scale-up SPEs~\cite{trill,lightsaber,saber,brisk,streambox,streamboxhbm,lifestream,grizzly} that follow hardware-conscious designs to maximize the single-machine performance of streaming query execution. State-of-the-art scale-up SPEs have shown that modern day multi-core machines with hundreds of processing cores and gigabytes of memory bandwidth are capable of meeting the performance demands of large scale streaming workloads and can be a cost-effective alternative to more expensive large multi-machine clusters~\cite{dowe}.
\section{Modern SPE Designs and Limitations}\label{sec:mot}
Scale-up SPEs attempt to achieve high single-machine performance primarily by three means. First, the user-defined query is subjected to several \emph{optimization} passes to prune redundant computations and increase data locality in order to improve the hardware utilization during query execution. Second, SPEs utilize different \emph{parallelization} strategies to take advantage of parallel processing cores available in multi-core machines. Finally, SPEs try to provide a \emph{low-overhead runtime} in order to meet the latency and throughput requirements of long running streaming applications. Despite the wide attention, we observe that the query optimization, parallelization, and execution strategies proposed in prior SPEs fail to generalize well on real-world streaming analytics applications due to several fundamental design limitations that we highlight below.

\noindent \textbf{Limited query optimization opportunities.} The optimization strategies adopted in current SPEs have limited applicability and often fail to cover a wide variety of real-world streaming applications. For instance, the majority of SPEs~\cite{struct, trill, streambox} adopt heuristics-based query optimization strategies, which are limited to basic graph transformations such as substituting or reordering individual operations in the query~\cite{struct, trill}. For instance, predicate pushdown~\cite{struct} is a common optimization where filtering operations (\emph{Where}) are moved closer to the data source in order to reduce the number of events the remaining operations in the query need to process. However, this optimization is only applicable if the predicate of the filtering operation is defined over the events in the input stream. For example, predicate pushdown cannot be applied to the example query in Figure~\ref{fig:q} as the \emph{Where} operation depends on the result generated by the parent \emph{Join} operation.

Certain advanced SPEs~\cite{lightsaber, grizzly} support more sophisticated low-level optimizations such as operator fusion for improving register/cache utilization by combining multiple operators into a single operator. For example, the \emph{Select} operators in the stock analysis query can be trivially fused with the \emph{Join} operator as shown in Figure~\ref{fig:fuseq}. This avoids unnecessary data movement between operators and allows intermediate results to remain in registers or cache memory for as long as possible. Unfortunately, the fusion rules implemented in current SPEs can only fuse operators until a so-called \emph{soft pipeline-breaker}~\cite{neumann, analyze} is reached. Soft pipeline-breakers are operators that require partial materialization of the output events before the next operator in the query pipeline can start processing. For instance, in Figure~\ref{fig:fuseq}, both \emph{Window-Sum} and \emph{Join} operators are soft pipeline-breakers. Fusing these operators together is non-trivial and optimizers in current SPEs fail in such scenarios. This significantly limits the applicability of fusion optimization on many real-world streaming applications as they often contain multiple pipeline-breakers in the query. For example, Table~\ref{tbl:apps} shows the temporal operations used in the queries of each application and each query contains between $2-6$ pipeline-breakers.

\noindent \textbf{Limited query parallelization capability.} Unlike batch processing applications, streaming queries do not inherently expose data parallelism as many temporal operations exhibit sequential data dependencies (e.g., sliding-window aggregation). Therefore, extracting data parallelism from streaming queries is often challenging and many SPEs rely on users to provide partitioned data streams in order to parallelize the query execution. For example, the trend analysis query (Figure~\ref{fig:q}) can be trivially parallelized by executing on data streams corresponding to different stocks. However, the degree of parallelism available from this approach is limited by the number of unique partitions available in the data stream~\cite{lightsaber}. Moreover, in certain streaming analytics applications used in healthcare~\cite{lifestream} and manufacturing industry~\cite{bearing}, even a single partition of the data stream can contain events generated at rates as high as $1-40$ KHz. In such cases, parallelizing the stream query execution requires more sophisticated strategies.

Prior works~\cite{lightsaber, saber, cutty, scotty} have proposed solutions to automatically extract data parallelism in streaming queries without needing a partitioned data stream. However, these works have been solely focusing on window-based aggregation operations. For example, the sliding-window sum in Figure~\ref{fig:ops}\textcolor{red}{c} can be parallelized by first computing partial sums on $5$-second tumbling-windows\footnote{Tumbling-window is a special case of sliding-windows when the stride length is same as the window size.} and then adding up two consecutive partial sums. Since the tumbling-windows do not overlap, the data streams can be partitioned on the $5$-second window boundaries and each window can be processed in parallel. The partial sum additions can also be parallelized through parallel reduction~\cite{lightsaber}. However, extending these strategies to parallelize arbitrary streaming queries is often non-trivial. For instance, determining the partition boundaries on stock price stream in the example query is unclear as the query contains multiple sliding-windows and a temporal join operation. We observe that the query parallelization methods in current scale-up SPEs~\cite{lightsaber, grizzly} are incapable of handling such scenarios.

\noindent \textbf{High runtime overhead during query execution.} The query execution model adopted in current SPEs fail to provide low overhead runtime for a wide range of real-world applications. The majority of the SPEs~\cite{trill, brisk, streambox, streamboxhbm, spark, flink} follow an interpretation-based query execution model also called an iterator model~\cite{iter, volcano}. In this model, the logical query description is translated to a data flow graph by mapping each temporal operator in the query to a concrete implementation. Each physical operator is designed to process events one-by-one or in micro-batches and passes the generated output events to the next operator in the graph through message queues. Despite being a widely adopted design, prior works~\cite{analyze, lifestream, grizzly, lightsaber, neumann, terse} have shown that interpreted SPEs often perform $1-2$ orders of magnitude slower compared to corresponding hand-tuned implementations. This inefficiency can be mainly attributed to the cost of data transfer between operators in the data flow graph~\cite{neumann, analyze}, poor support for effective optimization strategies such as operator fusion~\cite{analyze}, and failure to maintain end-to-end data locality because of fixed size micro-batching~\cite{lifestream}.

To address the inefficiencies of interpreted SPEs, recent works have proposed compiler-based solutions~\cite{lightsaber, saber, grizzly}, which can generate compact and efficient machine code from the high-level query description using customized code-generation techniques. Even though, compiler-based SPEs are shown to achieve state-of-the-art single-machine performance, we observe that these SPEs follow highly restrictive query languages with limited expressive power. To the best of our knowledge, prior compiler-based SPEs are designed only for queries performing window-based aggregation. Since it is common to use more complex and diverse set of temporal operations in streaming queries, the ability of current compiler-based SPEs to support real-world streaming analytics applications is significantly limited. Additionally, the code generation techniques used in these SPEs are primarily designed as source-to-source translator and heavily rely on template expanders. Such compiler designs are known to be highly inflexible and extremely hard to maintain~\cite{sysr, arch}.

Based on these observations, we conclude that the current SPEs only exploit the optimization and parallelization opportunities on stream queries in limited capacity due to several inherent design limitations. This prevents such SPEs from providing high-performance stream processing for many real-world streaming analytics applications. In this work, we set the goal to provide hardware-efficient stream processing without sacrificing programmability and generality to support the diverse computational requirements of modern day streaming analytics workloads.
\section{TiLT: A Time-Centric Approach}
Addressing the aforementioned design limitations and providing (i) effective query optimization, (ii) parallelization, and (iii) execution strategies on a diverse set of streaming analytics applications requires us to fundamentally rethink how SPEs should be designed. In this work, we argue that the limited query optimization capabilities and lack of automatic parallelization support stem from the event-centric temporal query representation model adopted in SPEs. Since the data streams are represented as a sequence of events, it is natural to define temporal operators as transformations over events. However, we observe that this event-centric definition of temporal operations do not fully express the time semantics of temporal queries necessary for effective query optimization and parallelization (see Section~\ref{sec:pass} for more details). Based on this observation, we make a fundamental shift from this established design principle and propose a new compiler-based SPE design that follows a \emph{time-centric} model for streaming queries.

As opposed to the traditional event-centric model, the time-centric model adopts a more fine-grained representation of temporal operations by defining them as functional transformations over well-defined time domains using a novel intermediate representation (IR) called \emph{TiLT}. TiLT IR is a highly expressive functional language with extensions to support temporal operations and offers several advantages over traditional query representation models. First, the functional definition of TiLT IR exposes inherent data parallelism which enables TiLT to parallelize \emph{arbitrary} streaming queries. Second, we demonstrate that the fine-grained time-centric definition of temporal operations allows TiLT to support effective query optimization strategies through simple IR transformations. Finally, we build a compiler-backend for TiLT that can automatically translate the time-centric IR query definitions to \emph{hardware-efficient} executable code.

\begin{figure*}
    \centering
    \includegraphics[width=\textwidth]{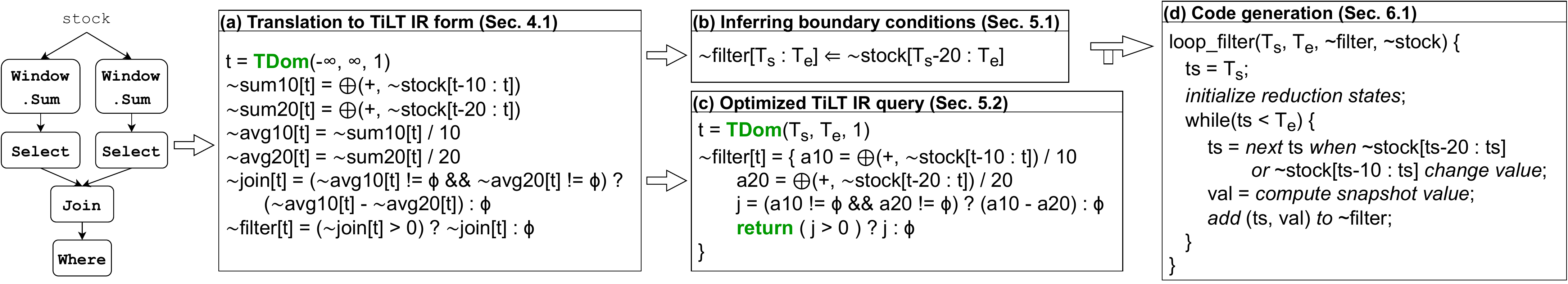}
    \caption{TiLT compilation pipeline}
    \label{fig:pipeline}
\end{figure*}

Figure~\ref{fig:pipeline} shows the lifecycle of a streaming application in TiLT. In the first stage, TiLT converts the streaming query written by the user into the TiLT IR form (Section~\ref{sec:tiltir}). After that, the compiler infers the boundary conditions necessary for parallelizing the query execution through a step called boundary resolution (Section~\ref{sec:bound}). Afterwards, TiLT queries are subjected to an optimization phase (Section~\ref{sec:fuse}). Finally, in the code generation step, the optimized query is lowered to LLVM IR and subsequently to executable code for parallel query execution (Section~\ref{sec:codegen}).

\subsection{TiLT IR}\label{sec:tiltir}
Similar to regular functional languages, TiLT supports data types such as integers, floating points, arrays, structures, dictionaries, and expressions such as arithmetic/logical operations, conditional operations, variables, and external function/library calls. On top of this, we introduce three new constructs, namely \emph{temporal objects}, \emph{reduction functions}, and \emph{temporal expressions}, which enables TiLT to define a diverse set of temporal operations. First, the temporal object is similar to a regular scalar object, except that it takes on a time-evolving value that spans over an infinitely long timeline. As opposed to the traditional way of representing streams as a sequence of discrete events with the time interval and payload fields, TiLT models the data stream using a single temporal object that assumes different values at different points in time based on the events active at that time. Second, TiLT uses the reduction functions to reduce the mutating values of a temporal object to a single scalar value. Reduction functions are introduced in TiLT to perform aggregate operations on data streams. Finally, temporal expressions are functional transformations on one or more temporal objects defined over a well-defined time domain. Temporal expressions are the basic building blocks of a streaming application in TiLT.

\noindent\textbf{Temporal object:} We use the $\sim$ notation to distinguish temporal objects from scalar objects. For example, let \tilt{stock} be a temporal object corresponding to a data stream of stock price events $e_i$ with price value $\rho_i$ and validity interval $(t_{i}^s, t_{i}^e]$. The value of \tilt{stock} at any point in time $T$ can be retrieved using an indexing operator ($[]$), and is defined as follows.

\begin{equation}
    \small
    \sim stock[T] = 
    \begin{cases}
        \rho_i    & \text{if } \quad \exists e_i \mid T \in (t_{i}^s, t_{i}^e] \\
        \phi        & \text{otherwise}
    \end{cases}
    \label{eqn:tobj}
\end{equation}

The value of the temporal object \tilt{stock} at time $T$ is the value of the payload of the stock price event active at $T$.\footnote{For simplicity, in this section, we assume that there are no events with overlapping intervals in the stream and therefore, there is only at most one event active at any given time. Handling data streams with overlapping events is discussed in Section~\ref{sec:codegen}.} When there are no events active at time $T$, the temporal object assumes a null value called $\phi$. The value $\phi$ has the special property that performing any arithmetic operations on it would always result in $\phi$. Additionally, TiLT allows defining derived temporal objects from existing ones by passing a time interval to the index. For example, the stock price values between time points $t_s$ and $t_e$ can be written as a derived temporal object \tilt{win} $=$ \tilt{stock}$[t_s:t_e]$. Then, the value of \tilt{win} at any point in time $T$ is defined as follows.

\begin{equation}
    \small
    \sim win[T] = 
    \begin{cases}
        \sim stock[T]    & \text{if } \quad T \in (t_s, t_e] \\
        \phi        & \text{otherwise}
    \end{cases}
\end{equation}

\noindent\textbf{Reduction function:} The reduce function, denoted as $\oplus \left(f, \sim I\right)$, is a special expression used to reduce a temporal object \tilt{I} into a scalar value based on a reduction operation $f$. TiLT, by default, supports several basic aggregation operations such as Sum ($+$), Product ($*$), Max ($>$), and Min ($<$). For example, reducing the temporal object \tilt{win} using summation ($+$) can be written as $\oplus \left(+, \sim win\right)$ and is defined as follows:
\vspace{-5pt}
\begin{equation}
    \oplus \left(+, \sim{win} \right) = \sum \{\sim{win}[t] \quad \forall t | \sim{win}[t] \neq \phi\}
\end{equation}

Other aggregation operations such as average can be expressed by combining the built-in reduction functions. On top of that, TiLT also allows users to define custom reduction operations (see Section~\ref{sec:codegen} for more details).

\noindent\textbf{Temporal expression:} TiLT expresses streaming queries as a sequence of temporal expressions each defining an output temporal object as a functional transformation over one or more input temporal objects on a time domain. A time domain \emph{TDom(start, end, precision)} has a start and end time indicating the interval between which the temporal expression is defined and a time precision denoting how frequently the value of the resulting temporal object can change in the time domain.
\begin{figure}[t!bhp]
    \vspace{-10pt}
    \includegraphics[width=0.8\columnwidth]{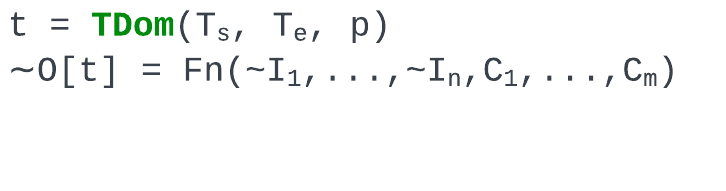}
    \vspace{-20pt}
    \label{fig:tilt_def}
\end{figure}

The general syntax of a temporal expression is shown above. The above expression defines the output temporal object \tilt{O} as a functional transformation ($Fn$) over $n$ input temporal objects \tilt{I_1}, ..., \tilt{I_n} and $m$ constants $C_1$, ..., $C_m$ over a time domain $t$. Here, $t$ is defined within the time interval $(T_s, T_e]$ and has a time precision of $p$. Therefore, at any point in time $T \in (T_s, T_e]$ and is a multiple of $p$, \tilt{O}$[T]$ assumes a value returned by the expression $Fn(\sim{I_1}, \sim{I_2}, ..., \sim{I_n})$ at time $T$.

\subsection{TiLT Queries}
\begin{figure}[t!bhp]
    \vspace{-10pt}
    \centering
    \includegraphics[width=.8\columnwidth]{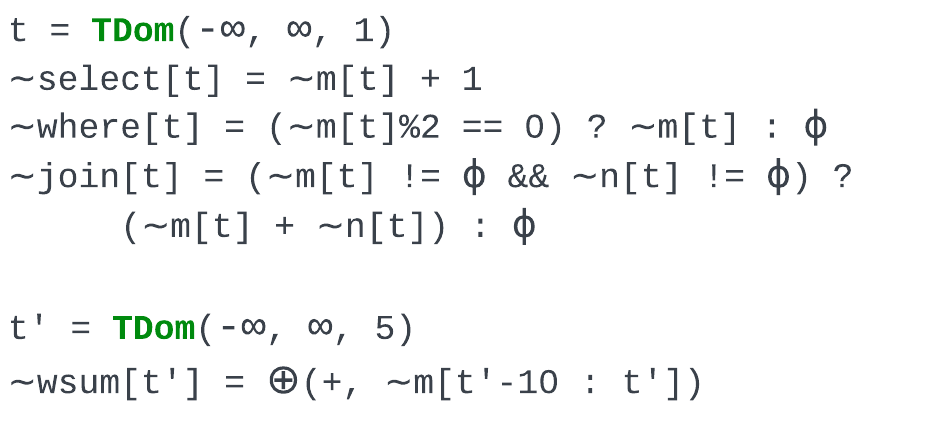}
    \vspace{-5pt}
    \caption{Temporal expression for \emph{Select}, \emph{Where}, \emph{Join}, and \emph{Window-Sum} operations}
    \vspace{-10pt}
    \label{fig:tiltq}
\end{figure}

Figure~\ref{fig:tiltq} shows the temporal expressions corresponding to the operations described in Figure~\ref{fig:ops}. In the example above, the time domain $t$ is defined between $-\infty$ and $+\infty$ and has a precision of $1$ second. That means the temporal expressions using $t$ define a value of the output temporal object at every second over an infinite time domain. The first temporal expression is equivalent to the \emph{Select} operation in Figure~\ref{fig:ops}\textcolor{red}{a}. This expression defines a functional transformation from the temporal object \tilt{m} to \tilt{select} over the time domain $t$ and the value of \tilt{select} at any point in time is defined to be $1$ more than the value of \tilt{m} at the same time. Similarly, the second expression is equivalent to the \emph{Where} operation and it filters only even values from \tilt{m}. In this example, the value of \tilt{where} at any point in time is conditionally selected to be $\phi$ if \tilt{m} has an odd value at that time. The third expression corresponds to the temporal join operation and follows a very similar structure to the \emph{Where} operation, except that it is a binary expression derived from two input temporal objects \tilt{m} and \tilt{n}. This expression identifies the strictly overlapping intervals between the events in \tilt{m} and \tilt{n} by checking if the both \tilt{m} and \tilt{n} have a non-null value at a given time. If yes, the expression returns the sum of the values, otherwise $\phi$. Finally, the $10$-second sliding-window \emph{Window-Sum} operation with $5$-second stride is defined by applying the \emph{Sum} reduction function on every $10$-second window derived from \tilt{m} over a time domain $t'$ with a precision of $5$. 

It should be noted that, even though the traditional definitions of these temporal operations have seemingly different semantics, TiLT definitions of the same operations are very similar in structure. This shows that TiLT is able to find a minimal set of abstractions necessary to represent a wide range of temporal operations. We believe that the integration of constructs such as temporal objects, reduction function, and temporal expressions into a powerful functional programming paradigm has made TiLT a highly expressive query representation model suitable for modern streaming applications.
\section{TiLT: Optimization and Parallelization}\label{sec:pass}
The compilation pipeline in TiLT starts by converting the streaming query into TiLT IR form as shown in Figure~\ref{fig:pipeline}\textcolor{red}{a}. Here, the example query is lowered to TiLT IR by defining the input data stream as a temporal object (\tilt{stock}) and mapping each operation to the corresponding temporal expression defined over an infinite time domain. After the translation, TiLT first extracts data parallelism from the query by resolving the boundary conditions of the TiLT expressions through a step called boundary resolution (Section~\ref{sec:bound}). Then, TiLT performs several optimization passes over the query through simple IR transformations. A full exploration of the optimization opportunities on TiLT IR is beyond the scope of this work. Instead, we focus on an optimization that provides the most bang-for-the-buck: operator fusion. Streaming queries generally exhibit high data locality and therefore can significantly benefit from operator fusion optimization that exploits this property to improve register/cache utilization~\cite{analyze, lifestream}. In the Section~\ref{sec:fuse}, we show how TiLT can overcome the limitations of current SPEs in performing operator fusion across pipeline-breakers.

\subsection{Boundary Resolution}\label{sec:bound}
The time-centric definition of the temporal queries in TiLT precisely captures the data dependency between temporal objects over the entire time domain. For example, the value of \tilt{join} at time $T$ in Figure~\ref{fig:tiltq} is only depended on the values of \tilt{m} and \tilt{n} at the same time. That means, the value of ~\tilt{join} at two different time points $T_1$ and $T_2$ are independent and can be evaluated in parallel. This data dependency information can be extended for the entire TiLT IR query. For example, the data dependency of \tilt{filter}$[T]$ in Figure~\ref{fig:pipeline}\textcolor{red}{a} can be determined by following the lineage all the way to the input temporal object \tilt{stock}. In this example, computing the value of \tilt{filter} at $T$ is solely dependent on the values of \tilt{stock} between time intervals $(T-10, T]$ and $(T-20, T]$. We call this the \emph{temporal lineage} of the temporal objects. TiLT uses this temporal lineage information to extract data parallelism from arbitrary temporal queries through a step called boundary resolution.

During the boundary resolution step, TiLT converts the initial query defined over the infinite time domain to a bounded domain by inferring the boundary conditions over the time domain. For example, based on the temporal lineage of the query in Figure~\ref{fig:pipeline}\textcolor{red}{a}, the values of \tilt{filter} between an arbitrary interval $(T_s, T_e]$ is only dependent on the values between the interval $(T_s-20, T_e]$ in \tilt{stock}. After the temporal boundary conditions have been inferred, TiLT redefines the time domain of the temporal query to the symbolic interval $(T_s, T_e]$ by setting $t$ to $TDom(T_s, T_e, 1)$ (Figure~\ref{fig:pipeline}\textcolor{red}{b}). TiLT uses this boundary condition to partition the data streams in order to parallelize the query execution (see Section~\ref{sec:exec} for more details).

\subsection{Operator Fusion}\label{sec:fuse}
Once the query is defined in the TiLT IR expression form, performing fusion optimization is straightforward through simple IR transformations. Applying operator fusion in TiLT queries entails simply merging two successive temporal expressions that are defined over the same time domain into a single expression. For example, following is the resulting expression after applying fusion rule on the temporal expressions \tilt{avg10}, \tilt{avg20}, and \tilt{join}.

\includegraphics[width=\columnwidth]{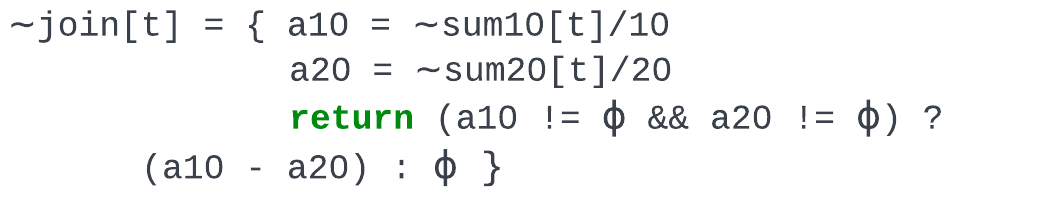}

Fusing expressions defined by \tilt{avg10}, \tilt{avg20}, and \tilt{join} simply requires replacing every occurrence of \tilt{avg10}$[t]$ and \tilt{avg20}$[t]$ in the \tilt{join} with \tilt{sum10}$[t]/10$ and \tilt{sum20}$[t]/20$ as shown above. This transformation is equivalent to the fusion optimization pass supported in current SPEs (shown in Figure~\ref{fig:fuseq}). However, unlike traditional SPEs, the same IR transformation can be applied to all the expressions in the query including the pipeline-breakers (e.g., \tilt{join}, \tilt{sum10}, \tilt{sum20}). TiLT repeatedly applies this transformation to fuse all temporal expressions in the trend-analysis query into a single expression as shown below.

\includegraphics[width=.9\columnwidth]{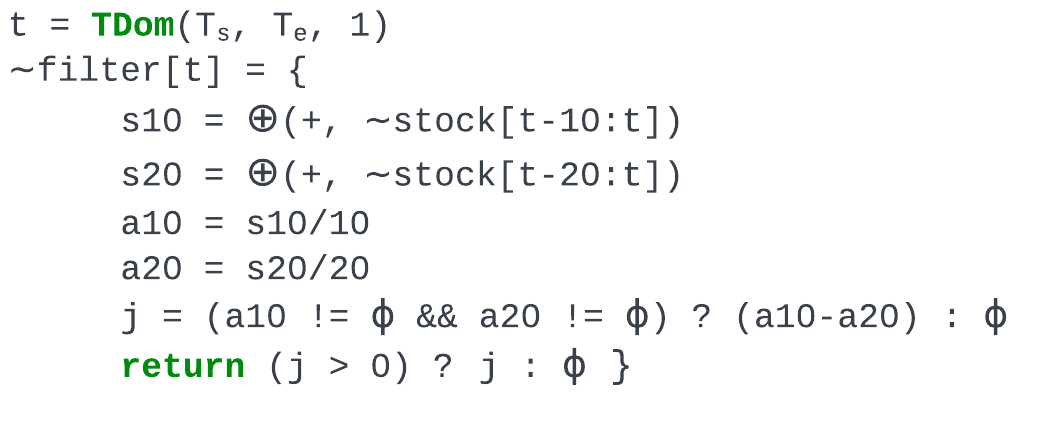}
\vspace{-5pt}

In comparison to current SPEs, TiLT supports more holistic and generalizable query optimization strategies because of the following two key reasons. First, the graph-level representation of streaming queries are typically too coarse-grained. Therefore, the fused version of the primitive operations is often not expressible at this level. TiLT, on the other hand, provides a more flexible query representation that allows fine-grained transformations as shown above. Second, the event-centric model often do not expose how the intervals of the events are manipulated after each temporal operator to the optimizer. For example, the intervals of the output events of \emph{Join} is only determined at runtime. In contrast, the time-centric definitions in TiLT IR expressions explicitly encode the transformations over time domains which allows TiLT to perform more sophisticated optimizations. We believe that TiLT IR opens up several more optimization opportunities on temporal queries that are otherwise hard to implement on traditional query representation models and we plan to explore them in the future work.
\section{TiLT: Compilation and Execution}\label{sec:codegen}
Even though the fine-grained compute definitions in TiLT IR is better suited for effective stream query optimizations, it also comes with significant amount of redundancy. For example, the temporal expression corresponding to the \emph{Select} operation shown in Figure~\ref{fig:tiltq} defines the value for the temporal object \tilt{select} at every second in the time domain. Although this fine-grained definition provides great flexibility for IR manipulation, it also introduces redundant computation since the value of the input temporal object \tilt{m} may not necessarily change every second. Therefore, na\"ively translating TiLT queries into executable code can potentially be highly inefficient. Below, we describe how TiLT compiler removes this redundancy and generates hardware-efficient executable code corresponding to TiLT IR queries.

\subsection{Code Generation}
TiLT compiler is written using C++ and LLVM JIT compiler infrastructure~\cite{llvm}. During the code generation phase, the compiler lowers the TiLT IR representation to LLVM IR. Since TiLT IR is fundamentally a functional language, it lends itself to standard code generation practices followed in compilers~\cite{advcomp, aho}. Below, we explain the code generation strategy used for the three newly introduced constructs in TiLT.
 
\subsubsection{Temporal Objects}
According to the formal definition in equation~\ref{eqn:tobj}, temporal objects define a value at every point in time. However, following the same definition for the physical implementation is impractical. Instead TiLT stores only changes in the value of the temporal object using a data structure called snapshot buffer (SSBuf). A snapshot buffer is an ordered sequence of snapshots stored in an array where each snapshot stores the timestamp ($ts$) and value ($val$) at the point when a change occurred.

\begin{figure}[h]
    \centering
    \vspace{-10pt}
    \includegraphics[width=.8\columnwidth]{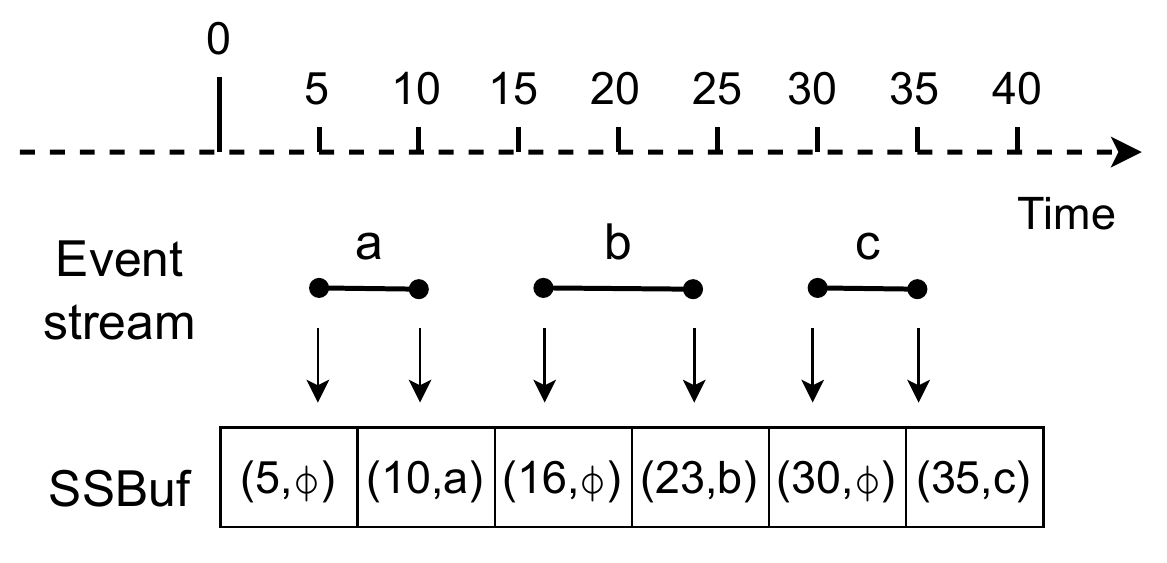}
    \caption{Event stream as snapshot buffer}
    \label{fig:snap}
\end{figure}

Figure~\ref{fig:snap} shows an example event stream stored as a snapshot buffer. The first snapshot in this buffer takes a value of null ($\phi$) at timestamp $5$ as there are no events active in the stream before that point. The second snapshot is added when first event ends (at $10$) and takes the value ($a$) of the payload of that event. Similarly, a new snapshot is added to the buffer at the start and end of every subsequent events in the data stream. When the data stream contains events with overlapping validity intervals, a single snapshot can assume multiple values. In such cases, TiLT uses a list/map to store the values of a snapshot.

\subsubsection{Reduction Functions}
We provide native support for several common reduction functions like Sum, Product, Min, and Max. Additionally, TiLT also supports user-defined reduction functions. Both built-in and user-defined functions are implemented using a general template similar to the ones followed in other SPEs~\cite{trill, lightsaber}. This template contains four lambda functions that is designed to incrementally update a state on every snapshot in the temporal object. (i) \emph{Init} function returns the initial state of the reduction operation (e.g., $0$ for Sum). (ii) \emph{Acc} function accumulates a single snapshot to the state (e.g., addition for Sum). (iii) \emph{Result} function returns the reduction result from the incremental state (e.g., the state for Sum). (iv) For invertible reduction functions, an optional \emph{Deacc} function can be provided that applies the inverse of the aggregate function on the state (e.g., subtraction for Sum). This simple template allows TiLT to support efficient aggregation implementations like Subtract-on-Evict~\cite{soe}.

\subsubsection{Temporal Expressions}
Finally, the temporal expressions are synthesized into loops that iterate over input snapshot buffers and update the output snapshot buffer. Figure~\ref{fig:pipeline}\textcolor{red}{d} shows the synthesized loop for the example query. The loop boundaries ($T_s$, $T_e$) and the loop counter ($ts$) increment is determined from the time domain boundaries and precision. The loop body performs the computation defined by the temporal expression. One iteration of the loop computes the snapshot value ($val$) of the output buffer \tilt{filter} at the timestamp $ts$.

However, as described above, na\"ively setting the loop counter increment based on time domain precision ($ts = ts+1$) might be highly inefficient as it introduces redundant iterations. Instead, TiLT takes advantage of an invariant of the functional definition of the temporal expressions to avoid redundant iterations, i.e., the output value of a temporal expression would only change when the inputs are changed. Based on this invariant, TiLT compiler generates an expression to increment the loop counter that computes the next value of $ts$ at which at least one of \tilt{stock[ts-10:ts]} and \tilt{stock[ts-20:ts]} have changed the enclosing snapshots. After loop synthesis, the generated loop is wrapped in a callable function with the symbolic loop boundaries parametrized as arguments (Figure~\ref{fig:pipeline}\textcolor{red}{d}). This allows TiLT to execute the query over any arbitrary time intervals on the output snapshot buffer.

\subsection{Query Execution}\label{sec:exec}
\begin{figure}[h]
    \centering
    \includegraphics[width=.9\columnwidth]{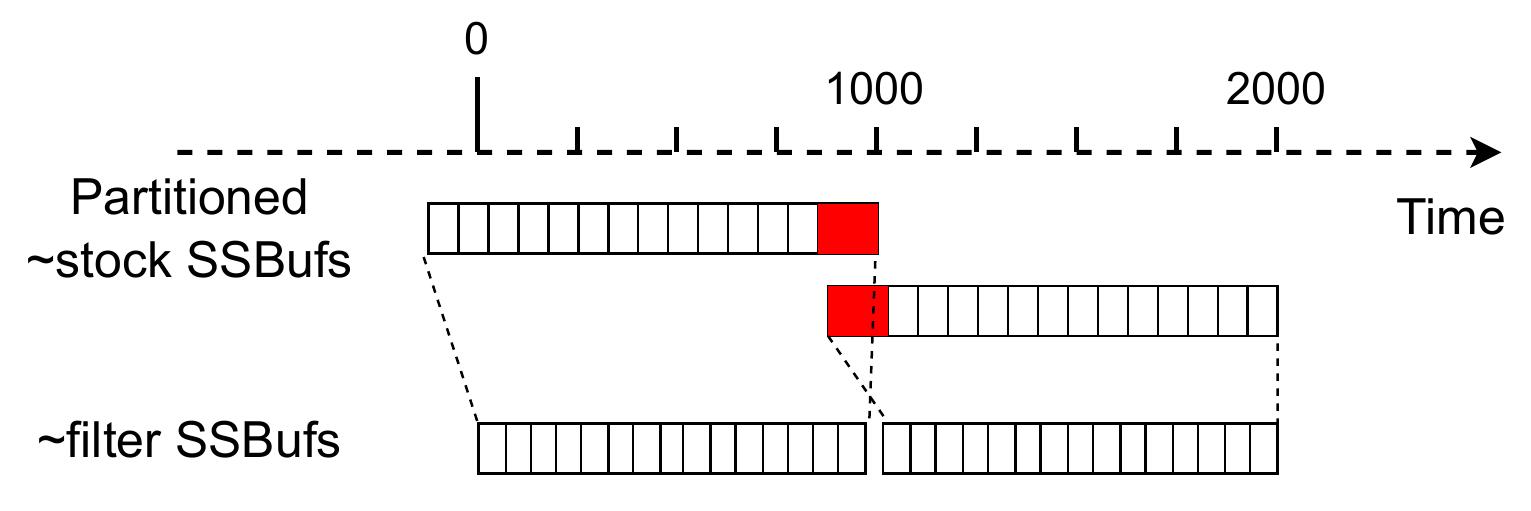}
    \caption{Parallel query execution}
    \label{fig:par}
\end{figure}

TiLT executes the generated code by partitioning the input data stream into snapshot buffers and processing them in parallel using independent worker threads. The data streams are partitioned based on the resolved boundary conditions (Figure~\ref{fig:pipeline}\textcolor{red}{b}) and a user-defined interval size. For example, Figure~\ref{fig:par} shows the partitioned snapshot buffers for the example query in Figure~\ref{fig:pipeline}\textcolor{red}{c} with an interval size of $1000$ seconds. In this example, computing the output snapshots in \tilt{filter} between the interval of $(0, 1000]$ requires reading snapshots from the input stream between the interval $(-20, 1000]$. Similarly, output snapshots between $(1000, 2000]$ require processing input snapshots between $(980, 2000]$ and so on. Even though, partitioning snapshot buffers as above adds some redundancy with duplicated snapshots (shaded area in Figure~\ref{fig:par}), extracting data parallelism like this allows executing continuous streaming queries using synchronization-free parallel worker threads.
\section{Evaluation}\label{sec:eval}
\noindent \textbf{Benchmarks: } For TiLT evaluation, we prepare two sets of benchmarks and a representative group of real-world streaming applications. (i) Temporal operations: this benchmark includes four commonly used primitive temporal operations shown in the Figure~\ref{fig:ops}. (ii) Yahoo streaming benchmark (YSB)~\cite{benchstream}: a popular streaming benchmark comprising a \emph{Select}, \emph{Where}, and a tumbling-window count operations. (iii) Real-world applications: this includes eight real-world streaming analytics applications shown in the Table~\ref{tbl:apps}.\footnote{We prepare these applications based on the realization that common benchmarks like YSB used for evaluating SPEs only represent a narrow set of real-world streaming analytics use-cases.} Table~\ref{tbl:apps} also describes the public datasets used in the evaluation. We provide more details on the benchmark queries in the Appendix~\ref{sec:bench}.

\noindent \textbf{Metrics: } Inline with prior works~\cite{spark, trill, lightsaber,streambox}, we use data processing \emph{throughput}, i.e., the number of events processed per second, as the primary comparison metric for the performance evaluation. Additionally, we also report latency-bounded throughput to evaluate the performance of different SPEs across a wide latency spectrum. Unless otherwise specified, the performance numbers are measured using a dataset with $160$ million events. All the numbers reported are the average measurements from $5$ runs of each experiment. The standard deviation of all the measurements is observed to be below $2\%$.

\noindent \textbf{Baselines: } On the temporal operations benchmarks, we compare the throughput of TiLT against four state-of-the-art scale-up stream query processing engines (SPEs): StreamBox~\cite{streambox}, Microsoft Trill~\cite{trill}, LightSaber~\cite{lightsaber}, and Grizzly~\cite{grizzly}. StreamBox and Microsoft Trill are both interpretation-based SPEs. StreamBox is written in C++ and uses pipeline parallelism to parallelize streaming queries. Trill is an SPE written in C\# designed to support diverse streaming analytics applications. Both LightSaber and Grizzly are compiler-based SPEs optimized for aggregation operations.

\noindent \textbf{Experimental setup: } All the experiments are conducted on AWS EC2 m5.8xlarge with 32 cores (with hyper-threading), $2.5$ GHz, and 128 GB DRAM. We also use AWS EC2 m5zn.3xlarge with 12 cores, $4.5$ GHz, and 48 GB DRAM for the scalability experiment. For a fair performance comparison, we exclude the time taken for disk and network accesses and only measure the compute performance of the query execution after loading the entire input dataset into the memory.

\subsection{Temporal Operations Throughput}
\begin{figure*}[t!bhp]%
    \centering
    \subfloat[Throughput on primitive temporal operations]{
        \includegraphics[width=0.5\textwidth]{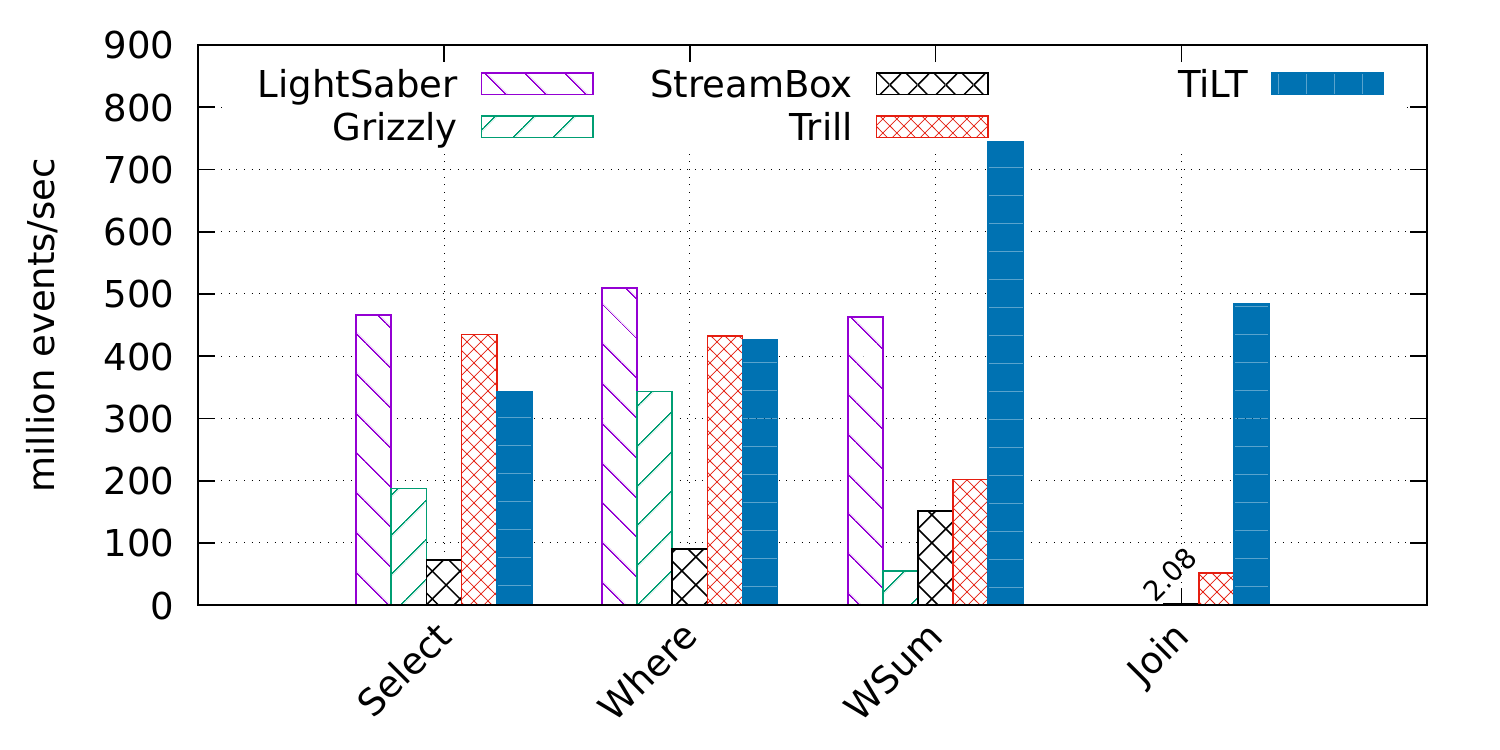}
        \label{fig:micro}
    }
    \subfloat[Throughput on real-world streaming applications]{
        \includegraphics[width=0.5\textwidth]{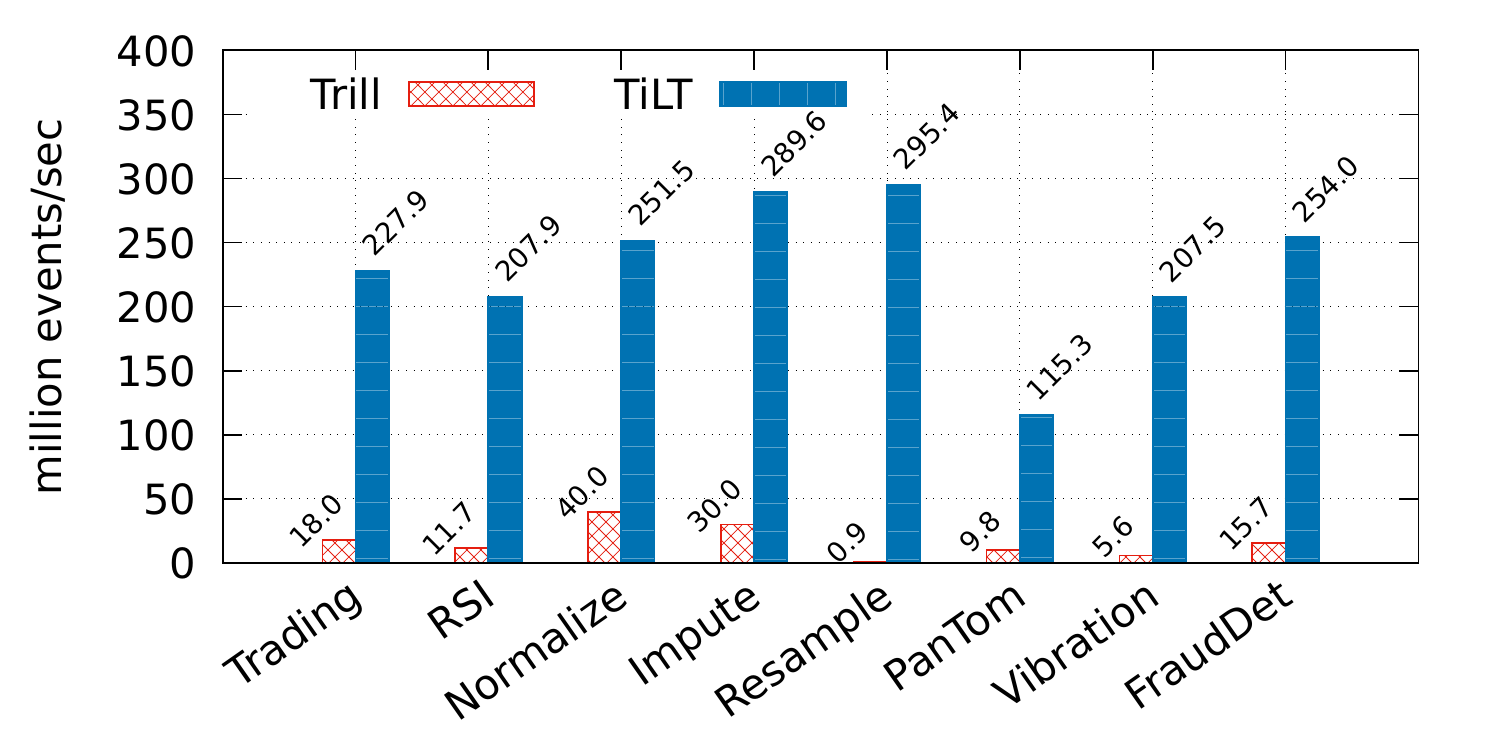}
        \label{fig:tp}
    }
    \caption{Performance comparison of TiLT on temporal operations and real-world streaming application benchmarks}%
\end{figure*}

We measure the performance of TiLT on the temporal operations \emph{Select}, \emph{Where}, \emph{Window-Sum}, and \emph{Join} on a synthetic dataset containing $160$ million events using 16 worker threads. Figure~\ref{fig:micro} shows the processing throughput comparison against StreamBox, Trill, Grizzly and LightSaber. For simple per-event operations like \emph{Select} and \emph{Where}, TiLT achieves similar performance to other SPEs (between $0.69-1.44\times$). On more complex operations like \emph{Window-Sum}, TILT outperforms the Trill and StreamBox by $6.64\times$ and $18.30\times$, respectively. This shows that TiLT generated code can significantly outperform hand-written operations in interpretation-based SPEs. Moreover, TiLT outperforms the two compiler-based SPEs Grizzly and LightSaber, which are optimized for window-based aggregation, by $7.44\times$ and $1.87\times$, respectively. We observe that the high overhead of Grizzly is caused by expensive atomic state updates used by the SPEs to perform parallel aggregation. LightSaber, on the other hand, uses complex data structures such as parallel aggregation tree which we observe to be inefficient for fine-grained window-aggregations that are common in modern streaming analytics applications.

Finally, we evaluate the performance of temporal \emph{Join}. Neither Grizzly nor LightSaber supports \emph{Join} operation, therefore we only compare the performance of TiLT against StreamBox and Trill. We observe that TiLT achieves $321.94\times$ higher performance over StreamBox and $13.87\times$ higher over Trill. The \emph{Join} operation in StreamBox is highly inefficient as it uses $O(n^2)$ algorithm to find overlapping events. Both Trill and TiLT follow in-order processing of the events and therefore only need $O(n)$ comparisons to perform the join. However, the Trill implementation uses expensive concurrent hashmaps to maintain operator states, whereas the time-centric model allows TiLT to generate more efficient state-free code for \emph{Join}. These results show that TiLT is able to generate highly efficient code for commonly used temporal operations that can significantly outperform both interpretation-based and compiler-based SPEs.

\begin{figure*}[t!bhp]%
    \centering
    \subfloat[Throughput on 12-core machine]{
        \includegraphics[width=0.5\textwidth]{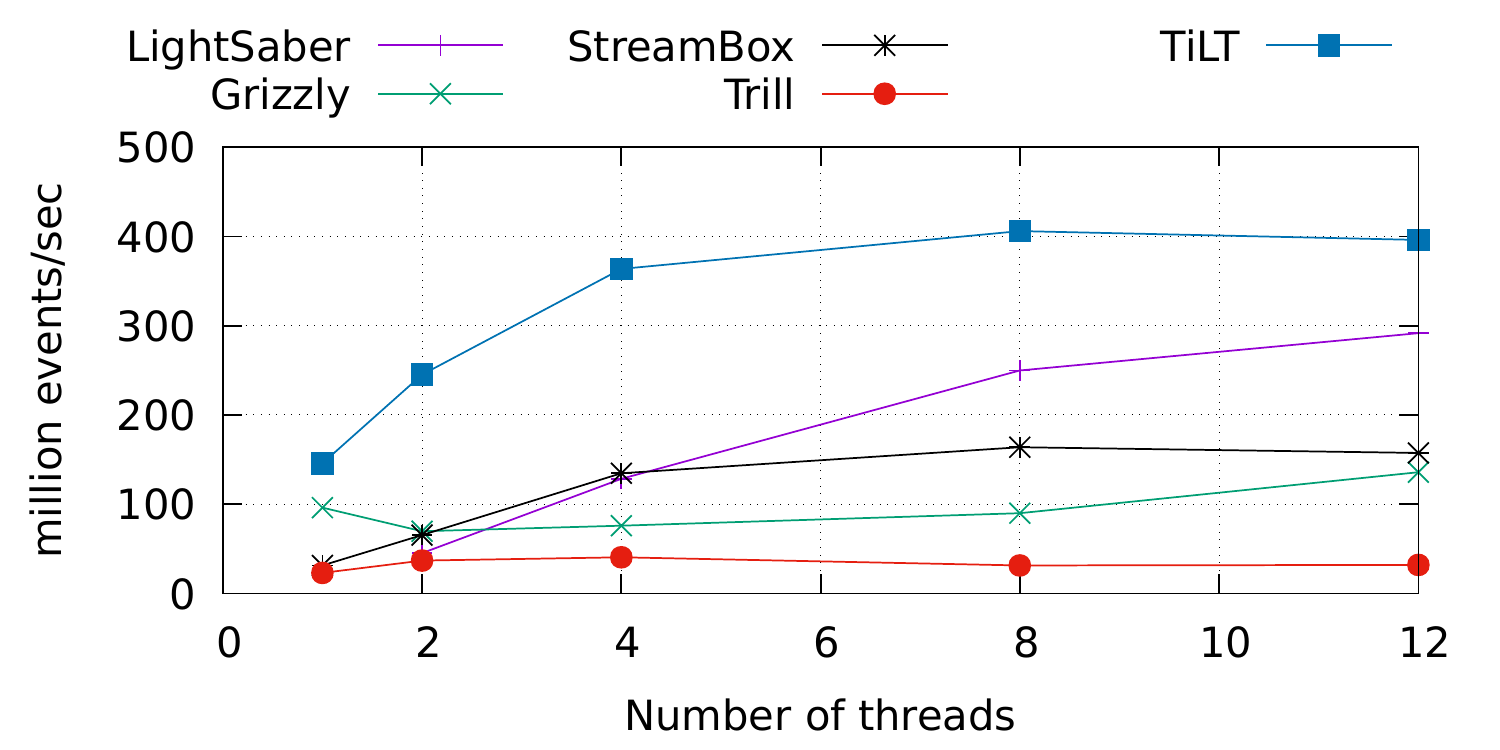}
        \label{fig:scale12}
    }
    \subfloat[Throughput on 32-core machine]{
        \includegraphics[width=0.5\textwidth]{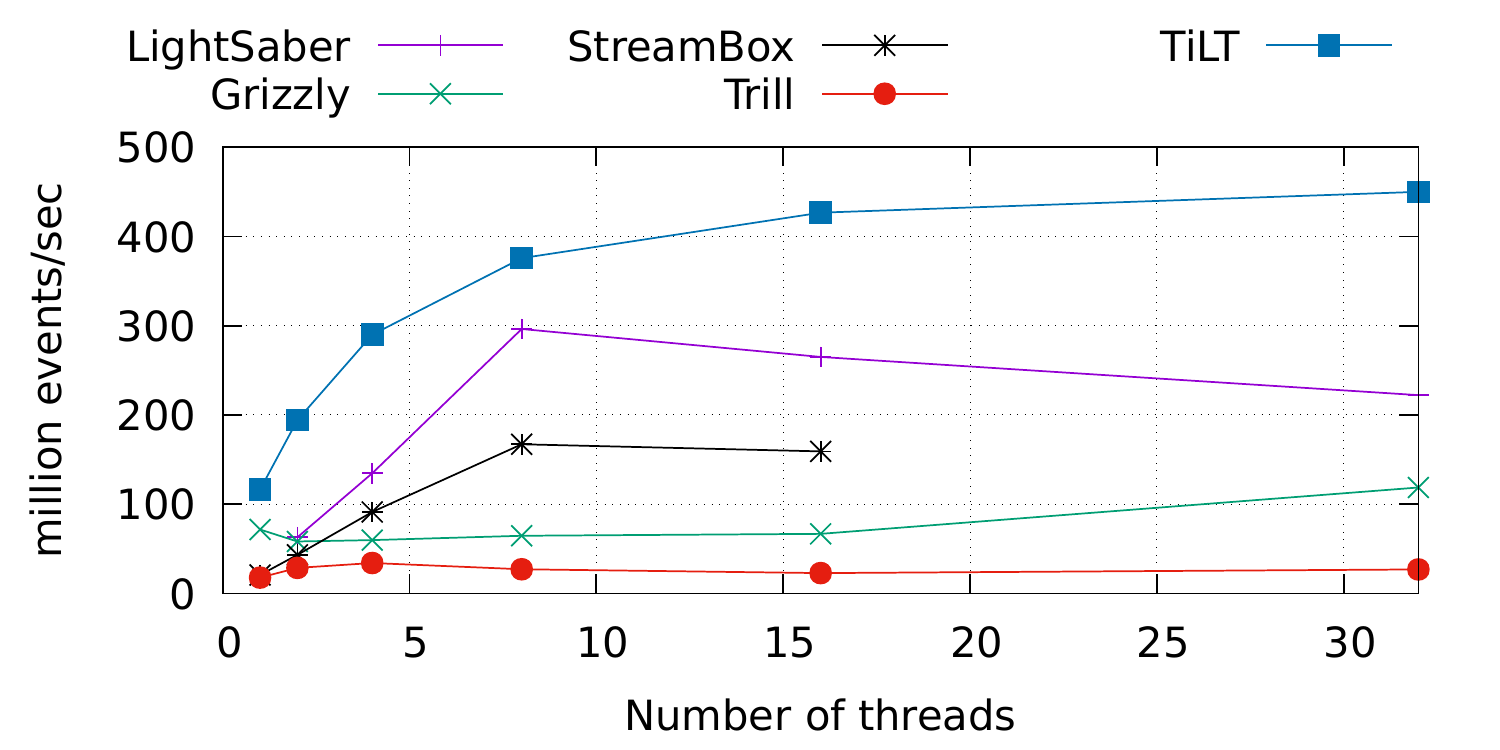}
        \label{fig:scale32}
    }
    \caption{Multi-core scalability on Yahoo Streaming Benchmark (YSB)}%
\end{figure*}

\begin{figure*}[t!bhp]
    \centering
    \includegraphics[width=\textwidth]{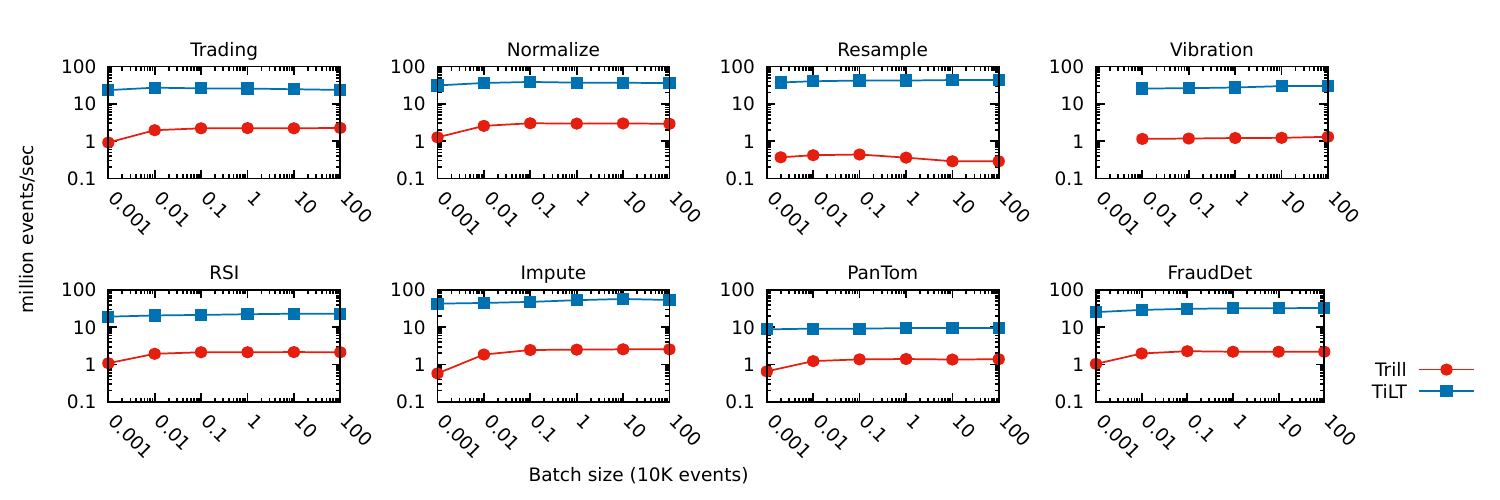}
    \caption{Latency-bounded throughput of Trill and TiLT}
    \label{fig:lat}
\end{figure*}

\subsection{Scalability}\label{sec:scale}
We evaluate how well TiLT can scale streaming queries over multi-cores using Yahoo Streaming Benchmark~\cite{benchstream}. We execute the query on a 12-core and a 32-core machine by increasing the number of worker threads and compare the throughput against Trill, StreamBox, Grizzly, and LightSaber. Figure~\ref{fig:scale12} and Figure~\ref{fig:scale32} show the throughput measured on the 12-core and 32-core machines, respectively. Trill only supports parallel execution over partitioned streams and exhibits the worst scalability. We also observe limited scalability in Grizzly. We believe this is due to concurrent data structures and atomic states used to synchronize between worker threads. Both StreamBox and LightSaber scale up to $8$ parallel threads on the 32-core machine. LightSaber achieves a peak multi-core performance of $291M$ events/sec and $296M$ events/sec on the 12-core and 32-core machine, respectively. TiLT consistently outperforms all the SPEs and achieves a peak performance of $406$ million events/sec on 12-core machine and $450$ million events/sec on 32-core machine. The superior performance of TiLT comes from the synchronization-free data parallel query execution strategy described in Section~\ref{sec:exec}. TiLT achieves close to linear scaling till 4-threads in the 12-core machine and 8-threads in the 32-core machine. The scalability benefits start to diminish afterwards because the query execution is shifting from being compute-bound to being memory-bound. This shows that TiLT can effectively parallelize the query execution while achieving $1.52-13.20\times$ higher peak performance over the state-of-the-art SPEs.

\subsection{Real-World Applications Performance}\label{sec:e2e}
We evaluate how well TiLT can support the performance requirements of real streaming workloads in comparison to state-of-the-art SPEs. To this end, we evaluate the throughput of TiLT on eight real-world streaming analytics applications listed in Table~\ref{tbl:apps}. These applications perform complex temporal transformations over data streams and therefore require a highly expressive temporal language for writing them as streaming queries. Out of all the baselines, we find that only Trill provides a query language that is capable of supporting all eight applications. Therefore, we compare the performance of TiLT on these applications against Trill.

First, we measure the throughput obtained from Trill and TiLT on these applications with 16 worker threads. As shown in Figure~\ref{fig:tp}, TiLT is able to outperform Trill across all the applications by $6.29-326.30\times$. This shows that TiLT is able to provide superior performance on a diverse set of streaming analytics applications. The best speedup is obtained on the signal resampling benchmark with $326.30\times$ higher throughput over Trill. This query requires a non-standard temporal operation called Chop, which we find to have an inefficient operator implementation in Trill. Despite this non-standard operation, TiLT is able to generate an efficient implementation that is ultimately resulted in a significant speedup.

Additionally, we also measure the latency-bounded throughput of TiLT against Trill on the real-world applications with the synthetic dataset. Trill is optimized to provide high throughput over a wide latency spectrum. As shown in Figure~\ref{fig:lat}, we measure the throughput by setting the batch/snapshot buffer size to contain events between $10$ and $1$M. We observe that TiLT provide consistently higher throughput across the entire latency spectrum, whereas Trill exhibits $18-227\times$ slowdown on smaller batch sizes due to high query execution overhead. This demonstrates that TiLT provides a runtime environment that adds minimal overhead and is able to provide high-performance over a wide latency spectrum.

\subsection{Effectiveness of Query Optimization}
\begin{figure}[h]
    \centering
    \includegraphics[width=\columnwidth]{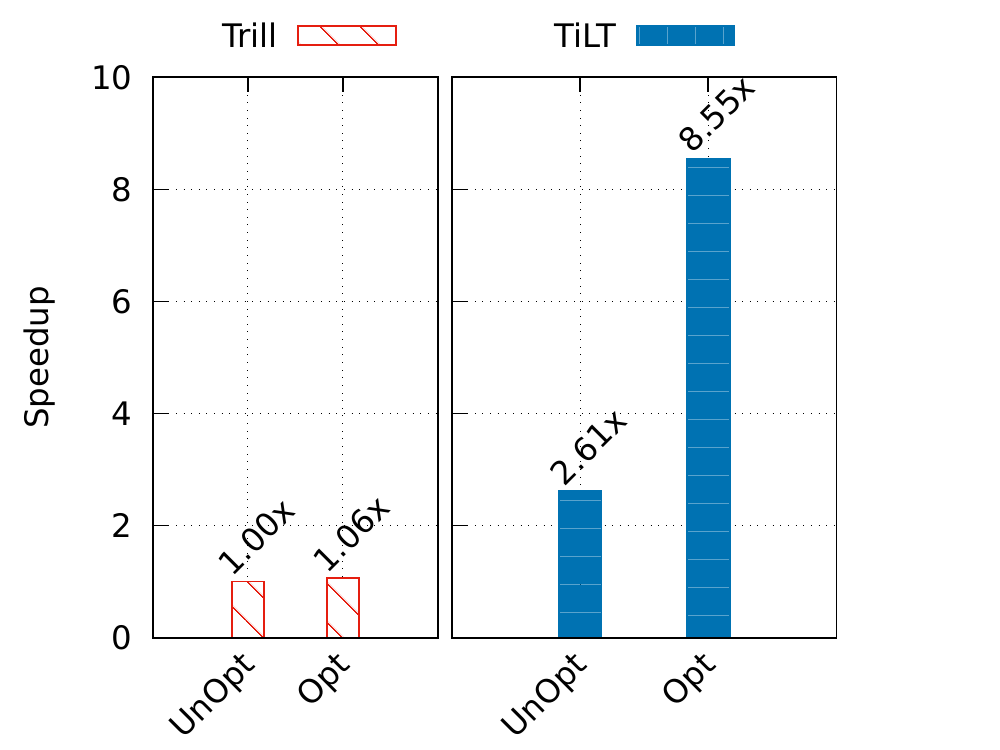}
    \caption{Performance breakdown of query optimization in Trill and TiLT (normalized to Trill)}
    \label{fig:fuse_perf}
\end{figure}

We analyze the effectiveness of the fusion optimization in TiLT by measuring the single-thread execution time of the example query (Figure~\ref{fig:pipeline}) before and after applying the IR transformations described in Section~\ref{sec:fuse}. We compare the result with the Trill version of the un-optimized and optimized queries shown in Figures \ref{fig:q} and \ref{fig:fuseq}. In Figure~\ref{fig:fuse_perf}, we report the speed up observed on each of these query versions normalized to the throughput of the un-optimized Trill query. As shown, applying operator fusion in Trill achieves only a nominal speedup of $1.06\times$. This highlights the limited optimization opportunities available in current scale-up SPEs. In TiLT, on the other hand, even the un-optimized version of the query outperforms the optimized Trill query by $2.61\times$. The TiLT query without applying the any optimizations (e.g., operator fusion) follows a similar query execution model as that of an interpreted SPE. Therefore, the speed up observed in this case can be mainly attributed to avoiding the common overheads associated with the managed language (C\#) implementation of Trill. This shows that TiLT can generate efficient code corresponding to individual operators that outperforms the hand-written implementations in interpreted SPEs. On top of this, the speedup can be further improved to $8.55\times$ after applying the operator fusion optimization. This speedup is the result of maximizing cache utilization by immediately reusing the intermediate results generated during query execution. This sensitivity study shows that the performance benefits of TiLT comes from both minimizing the runtime overhead using a compiler-based approach and by performing effective query optimizations enabled by the time-centric query representation model.

\section{Related Work}
Many of the stream processing engines (SPEs) commonly used in the industry (e.g., Apache Spark~\cite{spark}, Flink~\cite{flink}, Storm~\cite{storm}, Beam~\cite{dataflow}) are designed as scale-out systems based on the assumption that single machines are incapable to handle the performance demands of modern streaming analytics applications. However, recent works on scale-up SPEs~\cite{trill, streambox, streamboxhbm, lightsaber, saber, grizzly, brisk} have shown that a well-designed system running on a single multi-core machine can often satisfy the performance requirements of large scale stream processing~\cite{dowe}. Unfortunately, we observe that many of the state-of-the-art scale-up SPEs sacrifice the expressive power of the language to achieve high single-machine performance. In this work, we argue that to meet the demands of modern streaming analytics applications, it is important to strike a good balance between programmability and performance. Below, we compare state-of-the-art scale-up SPEs in terms of these two aspects.

Microsoft Trill~\cite{trill} is optimized for diverse streaming analytics applications and provides a highly expressive temporal query language with rich set of temporal operations and fine-grained windowing support. Trill uses efficient operator implementations and columnar data representation to maximize cache utilization and is shown to achieve $10-100\times$ higher single machine performance than scale-out SPEs~\cite{naiad}. However, Trill follows an interpretation-based query execution model that suffers from significant runtime overhead~\cite{analyze, neumann, lifestream}. TiLT, on the other hand, offers a similar query language but uses a compiler-based approach to generate hardware-efficient code that outperforms Trill by $20\times$ on average (Section~\ref{sec:e2e}). Other interpretation-based SPEs such as StreamBox~\cite{stream}, StreamBox-HBM~\cite{streamboxhbm}, BriskStream~\cite{brisk} are designed to achieve high performance on multi-core machines by efficiently parallelizing streaming queries. However, these SPEs only expose low-level APIs for writing streaming applications and offers limited temporal query language support. Moreover, we show that the synchronization-free data parallel query execution in TiLT achieve better scalability than these systems (Section~\ref{sec:scale}).

SABER~\cite{saber}, LightSaber~\cite{lightsaber}, Scabbard~\cite{scabbard}, and Grizzly~\cite{grizzly} are notable examples of recent compiler-based SPEs. However, as we explain in Section~\ref{sec:mot}, these systems are only optimized to efficiently execute the window-based aggregation and do not support important common temporal operations like temporal join. Therefore, these systems cannot support many real-world streaming analytics applications such as the ones shown in Table~\ref{tbl:apps}. TiLT, on the other hand, supports a wide range of streaming analytics applications with the help of the highly expressive temporal constructs in TiLT IR. We have also shown that TiLT is able to generate more efficient code and achieve superior performance compared to these SPEs (Section~\ref{sec:eval}). Moreover, the code generators in these SPEs are designed as template expanders which are harder to maintain and extend~\cite{arch}. In contrast, TiLT takes a more systematic approach by proposing a well-defined IR which we believe is important to facilitate further research in this field.
\section{Conclusion}
In this paper, we highlight the limitations of current stream processing engines (SPEs) and their inability to meet the performance demands of diverse set of modern day streaming analytics applications. To address these limitations, we design TiLT, a novel temporal query representation model for streaming applications. TiLT provides a rich programming interface to support a wide range of streaming analytics applications while enabling efficient query optimizations and parallelization strategies that are otherwise harder to perform on traditional SPEs. We also build a compiler-backend to generate hardware-efficient code from the TiLT query representation. We demonstrate that TiLT can outperform the state-of-the-art SPEs (e.g., Trill) by up to $326\times$ ($20.49\times$ on average) on eight real-world streaming analytics applications with diverse computational characteristics. TiLT source code is available at \href{https://github.com/ampersand-projects/tilt.git}{https://github.com/ampersand-projects/tilt.git}.

\section{Data-Availability Statement}
The artifact of this paper is published through Zenodo~\cite{artifact}.

\section*{Acknowledgments}
We first thank our shepherds and the anonymous reviewers for their valuable feedback and comments. We also like to thank members of the members of the EcoSystem lab, especially Kevin Song, Jasper Zhu, Xin Li, and Christina Giannoula for providing insightful comments and constructive feedback on the paper. This project was supported in part by the Canada Foundation for Innovation JELF grant, NSERC Discovery grant, AWS Machine Learning Research Award, and Facebook Faculty Research Award.

\appendix
\section{Real-world streaming applications}\label{sec:bench}
We prepare a benchmark suite with eight streaming analytics applications used in fields like stock trading, signal processing, industrial manufacturing, financial institutions, and healthcare. We prepare these applications based on the realization that commonly used benchmarking queries to evaluate stream processing engines (SPEs) like yahoo streaming benchmark (YSB)~\cite{benchstream} and Nexmark~\cite{grizzly} only represent a narrow set of real-world streaming analytics use-cases. Table~\ref{tbl:apps} provides a brief description of the streaming applications included in the benchmark suite and the corresponding public data sets used for the evaluation. In the following, we provide a detailed description of these applications. We also release the implementations of these queries in both Trill and TiLT as an artifact.

\noindent \textbf{Stock trading queries:} Streaming applications are widely used by investment services for analysing the trends in stock markets in order to make purchasing decisions. These applications continuously perform statistical algorithms on high-frequency stock price data streams. In our benchmark suite, we include two commonly used trading algorithms (i) Trend-based, and (ii) Relative strength index-based trading. The trend-based trading algorithm~\cite{invest} computes short-term and longer-term moving averages (e.g., $20$ minutes and $50$ minutes) of each stock price over time and identifies an upward trend when the short-term average goes above long-term and vice versa for the downward trend. The second trading algorithm uses relative strength index (RSI)~\cite{rsi} as the momentum indicator instead of the moving averages. RSI is an indicator to chart the current and historical strength or weakness of a stock or market based on the closing prices during a $14$-day trading period. These two algorithms are widely used and are often combined with more sophisticated trading algorithms.

\noindent \textbf{Data cleaning/preparation:} The real-time data processed by streaming applications are usually misformatted, corrupt and garbled. Therefore, data analysts often need to conduct data cleaning and preprocessing before analysing the raw data streams. For example, events collected from different sources often vary widely in their scale of values. Data normalization is a commonly used approach to bring the values of the events on different scales to a notionally common scale. We include a standard score-based normalization query which computes the mean ($\mu$) and standard deviation ($\sigma$) of the event payload values ($X$) over every $10$-second tumbling window. The values of each event in the window is normalized by computing $(X-\mu)/\sigma$.

Secondly, signal processing operations are used in the healthcare industry to clean and prepare the physiological signals like ECG and EEG collected from the patients~\cite{lifestream}. These signals are collected at a fixed frequency usually ranging from $10^{-4}$ Hz to $10^3$ Hz. We include two commonly used signal processing operations in our benchmark suite: (i) signal imputation, and (ii) signal resampling. Signal imputation operations are used to fill missing events in the signal streams. The naive imputation approaches include substituting the missing signal values with a constant (e.g., zero) or with the value of the last active event. We include a signal imputation query that replaces the missing signal values with the average values of the events in their corresponding $10$-second tumbling window. The signal resampling operation is used to translate a signal stream in one frequency to another. We use the linear-interpolation\cite{resample} algorithm to perform the frequency conversion on the signal streams.

Finally, PanTomkins algorithm~\cite{pantom} is commonly used to detect QRS complexes in ECG signals. The QRS complex represents the ventricular depolarization and the main spike visible in an ECG signal and is used to measure the heart rate of the patients.

\noindent \textbf{Manufacturing industry:} Industrial sensors used for monitoring the health of large machinery in the manufacturing industry. One such use case is monitoring the vibration signals of the ball bearings in order to predict their failure rates. These sensors often generate data streams at a very high frequency (as much as $40$ KHz frequency)~\cite{bearing} and the standard vibration analysis algorithms require computing complex aggregate functions on such streams. We include a vibration analysis query that computes three window-based aggregate functions over the data stream, namely, kurtosis~\cite{kurt}, root mean square~\cite{rms}, and crest factor~\cite{crest} over a $100$-millisecond tumbling window.

\noindent \textbf{Financial institutions:} It is important for the financial and banking institutions to identify fraudulent activities by analyzing the real-time financial transaction data of their customers. One of the basic fraud detection strategy is a rule against abnormal transaction quantities. The fraud detection query in our benchmark suite computes a moving average ($\mu$) and standard deviation ($\sigma$) on the purchasing quantity on the transactions for each individual over a $10$-day sliding window and calculates the threshold for large quantity as $\mu + 3 * \sigma$. Finally, a filtering operation is applied to select the transactions that crosses the large quantity threshold and marks as a potential fraudulent transaction.
%
%
%
%
%


\section{Artifact Appendix}\label{sec:ae}

\subsection{Abstract}
We provide the source code and scripts to reproduce the scalability results (in Section~\ref{sec:scale}) and the real-world applications performance (in Section~\ref{sec:e2e}) in the main paper. This appendix contains instructions to generate plots similar to Figure~\ref{fig:scale12}, Figure~\ref{fig:scale32}, and Figure~\ref{fig:tp} on synthetically generated data sets. The performance numbers measured on synthetic data set should be a close estimate of the same on the real data set.

We include docker containers to setup the runtime environment for all the experiments in order to support portability. Therefore, the artifact can be executed on any multi-core machine with docker engine installed. We also use Linux gnuplot utility to generate figures from the collected performance numbers. We \textbf{recommend} using Ubuntu 20.04 operating system for running the scripts provided in the artifact.

\subsection{Artifact Checklist}

{\small
\begin{itemize}
  \item {\bf Algorithm:} Not applicable.
  \item {\bf Program:} Benchmarks described in the Table~\ref{tbl:apps}.
  \item {\bf Compilation:} Provided as dockerized containers.
  \item {\bf Transformations:} No transformation tools required.
  \item {\bf Binary:} Source code and scripts included.
  \item {\bf Data set:} A synthetic data set is provided along with the artifact.
  \item {\bf Run-time environment:} Docker files provided to create runtime environment.
  \item {\bf Hardware:} A single multi-core CPU. Ideally, with at least $16$ cores and $128$ GB memory.
  \item {\bf Runtime state:} Not sensitive to runtime state.
  \item {\bf Execution:} Less than an hour to evaluate all the benchmarks.
  \item {\bf Metrics:} Number of events processed per second (Throughput).
  \item {\bf Output:} Plots similar to Figure~\ref{fig:scale12}, Figure~\ref{fig:scale32}, and Figure~\ref{fig:tp} in the main paper.
  \item {\bf Experiments:} Bash scripts and docker files are provided to run the benchmarks. Numerical variations in the results are negligible.
  \item {\bf How much disk space required (approximately)?:} $\sim50$ GB.
  \item {\bf How much time is needed to prepare workflow (approximately)?:} Under $1$ hour to setup the runtime environments.
  \item {\bf How much time is needed to complete experiments (approximately)?:} Under $30$ minutes.
  \item {\bf Publicly available?:} Yes
  \item {\bf Code licenses (if publicly available)?:} LGPL-3.0
  \item {\bf Data licenses (if publicly available)?:} Not applicable.
  \item {\bf Workflow framework used?:} No.
  \item {\bf Archival link:} \DOI
\end{itemize}
}

\subsection{Description}

\subsubsection{How to Access}
The artifact can be downloaded either from the GitHub link \github or from the DOI link \DOI.

\subsubsection{Hardware Dependencies}
TiLT does not require any special hardware. A single general purpose multi-core CPU should be sufficient for running the artifact. We recommend to use a machine with at least $16$-cores and $128$ GB memory.

\subsubsection{Software Dependencies}
The experiments provided in this artifact is prepared to run inside a docker container. We recommend to use a machine with Ubuntu 20.04 with docker installed to reproduce the results. Additionally, we use gnuplot generate plots/figures from the performance numbers.

\subsubsection{Benchmarks and Baselines}
The experiments provided in this artifact uses the Yahoo Streaming Benchmark (YSB)~\cite{benchstream} to perform the scalability experiments described in Section~\ref{sec:scale} and streaming applications shown in Table~\ref{tbl:apps} to measure real-world application performance described in Section~\ref{sec:e2e}. We include scripts to build and run these experiments on Trill~\cite{trill}, StreamBox~\cite{streambox}, Grizzly~\cite{grizzly}, LightSaber~\cite{lightsaber}, and TiLT. We use query processing throughput as the comparison metric for all the benchmarks, i.e., the number of events processed per second.

\subsubsection{Data Sets}
For convenience, we provide a synthetically generated data set for all the experiments. The results produced on the synthetic data set should be comparable to the results we report on the real data set in the main paper.

\subsection{Installation}
We provide docker files to setup the runtime environment for all the experiment.
\begin{enumerate}
    \item Install docker following the instructions in\\ \href{https://docs.docker.com/engine/install/ubuntu/}{https://docs.docker.com/engine/install/ubuntu/}.
    \item Install gnuplot by running the following command:\\
    \texttt{\small sudo apt-get install -y gnuplot}
    \item Clone the git repository using the following command: \\
    \texttt{\small git clone \\     https://github.com/ampersand-projects/streambench.git \\     {-}{-}recursive}
    \item Build the docker images by running the \texttt{setup.sh} script at the root directory of the cloned repository.
\end{enumerate}

\subsection{Experiment Workflow}
Execute \texttt{run.sh} to run all the experiments and generate the figures.

\subsection{Evaluation and Expected Results}
Once the script has finished execution, they would have generated plots similar to Figure~\ref{fig:scale12} and \ref{fig:tp}. The figures can be found at the root directory of the repository under the names \texttt{ysb.pdf} and \texttt{e2e.pdf}.

First, \texttt{ysb.pdf} plots the throughput comparison of TiLT against Trill, StreamBox, Grizzly, and LightSaber on the Yahoo Streaming Benchmark (YSB)~\cite{benchstream} on different degree of parallelism ranging from $1$ to $16$. Compared to other baselines, TiLT should consistently achieve higher throughput and scalability. Second, \texttt{e2e.pdf} plots the throughput comparison of TiLT against Trill on the real-world streaming applications listed in Table~\ref{tbl:apps} both using a fixed parallelism of $8$ threads. On an average, TiLT should achieve $\sim 10-100\times$ higher throughput compared to Trill.

\subsection{Experiment Customization}
The parallelism for the real-world applications performance evaluation experiment can be modified by setting the \texttt{\$THREADS} environmental variable to appropriate number of threads in the scripts \texttt{trill\_bench/run.sh} and \texttt{tilt\_bench/run.sh}.

\subsection{Methodology}

Submission, reviewing and badging methodology:

\begin{itemize}
  \small
  \item \url{https://www.acm.org/publications/policies/artifact-review-badging}
  \item \url{http://cTuning.org/ae/submission-20201122.html}
  \item \url{http://cTuning.org/ae/reviewing-20201122.html}
\end{itemize}



\end{document}